\documentclass[11pt]{article}
\usepackage[margin=1in]{geometry}
\usepackage{adjustbox}

\usepackage[utf8]{inputenc} 
\usepackage[T1]{fontenc}    
\usepackage{graphicx}
\usepackage{subfigure}

\usepackage{amsmath}
\usepackage{amssymb}
\usepackage{dsfont}
\usepackage{mathrsfs}
\usepackage{hyperref}
\usepackage{braket}
\usepackage{bbold}
\usepackage{cancel}
\usepackage{units}
\usepackage{array}
\usepackage{enumitem}
\usepackage{url}            
\usepackage{booktabs}       
\usepackage{amsfonts}       
\usepackage{nicefrac}       
\usepackage{microtype}   
\usepackage{hyperref}
\usepackage{algorithm2e}
\usepackage{multirow}
\usepackage{multicol}
\usepackage{todonotes}
\usepackage{authblk}

\DeclareMathOperator*{\argmin}{arg\,min}

\newcommand{\Hb}{\boldsymbol{H}}

\newcommand{\Ub}{\boldsymbol{U}}

\newcommand{\Xb}{\boldsymbol{X}}

\renewenvironment{abstract}
{\centerline{\large\bf Abstract}\vspace{0.7ex}%
  \bgroup\leftskip 20pt\rightskip 20pt\small\noindent\ignorespaces}%
{\par\egroup\vskip 0.25ex}

\usepackage{booktabs, multirow, ltablex}

\newenvironment{keywords}
{\bgroup\leftskip 20pt\rightskip 20pt \small\noindent{\bf Keywords:} }%
{\par\egroup\vskip 0.25ex}
\newlength\aftertitskip     \newlength\beforetitskip
\newlength\interauthorskip  \newlength\aftermaketitskip

\newcommand{\BlackBox}{\rule{1.5ex}{1.5ex}}  
\ifdefined\proof
    \renewenvironment{proof}{\par\noindent{\bf Proof\ }}{\hfill\BlackBox\\[2mm]}
\else
    \newenvironment{proof}{\par\noindent{\bf Proof\ }}{\hfill\BlackBox\\[2mm]}
\fi
 
\newtheorem{theorem}{Theorem}
\newtheorem{lemma}[theorem]{Lemma} 
\newtheorem{proposition}[theorem]{Proposition}

\RequirePackage{epsfig}
\RequirePackage{amssymb}
\RequirePackage{natbib}
\RequirePackage{graphicx}

\if@usehyper
\RequirePackage[colorlinks=false,allbordercolors={1 1 1}]{hyperref}
\fi

\if@abbrvbib
\else
\fi

\bibpunct{(}{)}{;}{a}{,}{,}

\begin{document}

\title{Identifying Invariant Factors Across Multiple Environments with Kullback-Leibler Regression}

\author[1]{Jaime Roquero Gimenez}
\author[2]{James Zou}
\date{}
\affil[1]{Department of Statistics, Stanford University}
\affil[2]{Department of Biomedical Data Science, Stanford University}

\maketitle

\begin{abstract}
Many datasets are collected from multiple environments (e.g. different labs, perturbations, etc.), and it is often advantageous to learn models and relations that are invariant across environments. Invariance can improve robustness to unknown confounders and improve generalization to new domains. We develop a novel framework that we term Kullback-Leibler regression (KL regression) to reliably estimate regression coefficients in a challenging multi-environment setting, where latent confounders affect the data from each environment. KL regression is based on a new objective of simultaneously minimizing the sum of Kullback-Leibler divergences between a parametric model and the observed data in each environment, and we derive an analytic solution for its global optimum. We prove that KL regression recovers the true invariant factors under a flexible confounding setup. Extensive experiments show that KL regression performed better than state-of-the-art causal inference techniques across a variety of settings, even with large model mismatch. Moreover KL regression  achieved the top score on a DREAM5 challenge for inferring causal genes.  
\end{abstract}


\section{Introduction}

Drawing conclusions by analyzing data from a single source carries serious challenges. For example, if a gene expression-phenotype correlation is found to be statistically significant in one lab, does it imply that we will be able to replicate it in other settings? It is possible that some shared latent characteristic between the samples explains such correlation, rather than a true biological connection between the gene and the disease. Confounding is a pervasive problem, where unobserved variables that change across environments invalidate any result made on a single dataset~\citep{zou2014epigenome}. Leveraging multiple data sources---also called environments---can help to identify the relationship between variables that are common across diverse settings and hence are more likely to be unconfounded true relations~\citep{ben2010theory}. However, how best to integrate multiple environments and when would it work are very much open areas of research. This question has recently generated wide interest in the machine learning community due to its connections to causal inference and robustness to domain shift~\citep{ben2010theory,imbens2015causal,kouw2016feature}. Environments can correspond to different batches, labs where the experiment is done, perturbations, etc. One approach is to first fit a regression model on each environment separately to obtain the environment $e$ specific parameters $\beta_e$ between the covariates and the response. Then some meta-analysis can be performed to aggregate the different $\beta_e$, or at the level of p-values if those are available. In the presence of unknown confounders, however, such two-stage meta-analysis can produce erroneous results. Other models introduce the environment as part of the model, and fit all the data where the heterogeneity across environments is captured by methods such as mixed-effects models \citep{pinheiro2000linear}, time varying coefficients \citep{hastie1993varying,fan1999statistical}, etc. However, these models assume a structured transformation across environments, and cannot capture a potential radical change in the covariate structure from one environment to another due to an experimental manipulation. Another approach is to focus on the invariant element of the model that we want to identify, and look for such invariant signal across environments \citep{peters2016causal}. We follow this line of work, where techniques differ from each other through their scope of application and their goal: some estimate the invariant vector of coefficients, while others provide a vector of coefficients that solves a minimax risk problem over the different environments. We will further discuss the relevant literature at the end of Section~\ref{section:section-2}.
\paragraph{Our Contributions.} We present a novel method for identifying invariant factors across multiple environments that we call \emph{Kullback-Leibler regression} (KL regression) focusing on identifying the covariates that have a causal effect on the response. Unlike other approaches, KL regression simultaneously allows for the existence of \emph{latent confounders}, is valid in settings where the different environments arise from distribution shifts in the observed covariates \emph{and} shifts in the response, and \emph{provably recovers the invariant causal vector} of coefficients. We provide asymptotic convergence results and demonstrate its improved performance compared to standard approaches in extensive experiments. Additionally, we introduce a \emph{Lasso KL regression} estimator, where a sparsity inducing penalty is added to the problem defining our estimator. In practice, this variant under a sparsity assumption on $\beta^*$ shows remarkable stability and robustness to model misspecification. We begin in Section~\ref{section:section-2} by presenting the problem of linear regression estimation with confounding due to latent variables based on linear Structural Equation Models (SEM) and describing the relevant literature. In Section~\ref{section:section-3} we describe KL regression in its general form, then present the required parametrization steps tailored to our SEM model that lead to our KL regression estimator (both sparse and non-sparse versions). We thoroughly validate the theoretical results empirically in Section~\ref{section:section-4} and compare our method to other commonly used techniques. KL regression achieves the top score for a DREAM5 challenge.

\section{Confounding with Structural Equation Models}\label{section:section-2}

\subsection{Generating Multiple Environments Via Linear Structural Equation Models}

\begin{figure}
    \centering
    \includegraphics[width=.4\linewidth]{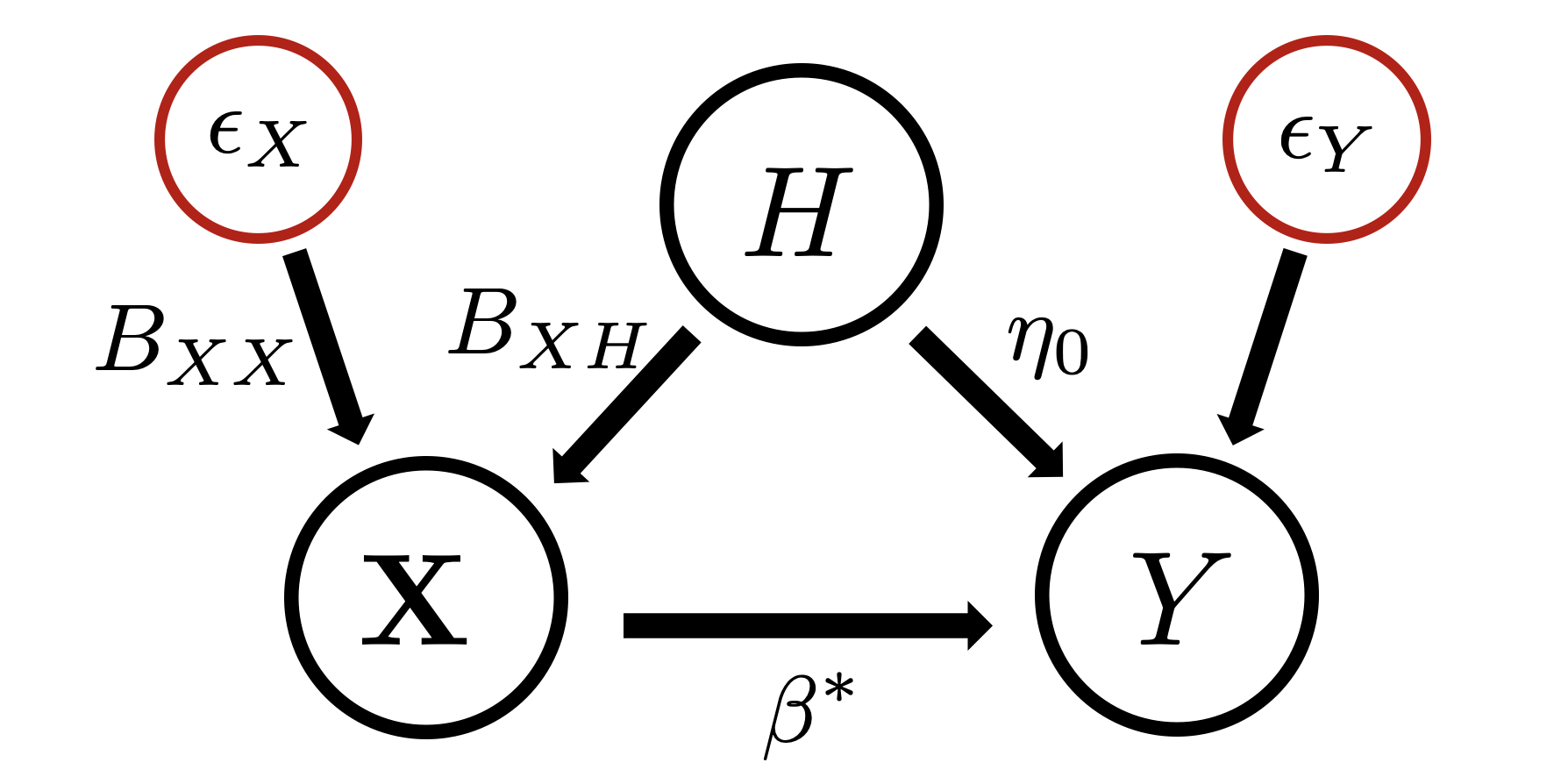}
    \caption{Graphical representation of the process generating multiple environments.}
    \label{fig:DAG}
\end{figure}

Latent variables that introduce a confounding effect on the observed data is a common framework to represent spurious effects. In a linear setting, we may assume that the unconfounded vector of coefficients---the target of interest---which relates the response to the observed covariates represents an invariant structure that has a causal interpretation \citep{rojas2018invariant}. However, estimators like least squares regression may be biased due to the presence of latent confounding effects. We consider the case where the data follows a linear Structural Equation Model (SEM) \citep{bollen1989structural,kaplan2008structural}, which defines a distribution $\pi$ over the observed covariates $\Xb\in \mathbb{R}^D$, the hidden covariates $\Hb\in \mathbb{R}^Q$ and the observed response $Y\in \mathbb{R}$ as a solution to the system of equations:
\begin{align}
    \begin{cases}{}\label{equation-linear-sem-individual}
    H_q = \epsilon_{H,q} \quad \text{for $1\leq q \leq Q$}
    \\ X_d  = \sum_{1\leq d' \leq D} [B_{XX}]_{dd'}X_{d'}\! +\! \sum_{1\leq q\leq Q}[B_{XH}]_{dq}H_q 
    \\ \hspace{2cm} + \; \epsilon_{X,d} \qquad \text{for $1\leq d \leq D$}
    \\ Y = \sum_{1\leq d \leq D} \beta^*_{d}X_d + \sum_{1\leq q \leq Q} \eta_{0,q} H_q + \epsilon_Y
    \end{cases}
\end{align}{}
where the coefficients are given by the \emph{connectivity} matrices $B_{XX}\in\mathbb{R}^{D\times D}$, $B_{XH}\in\mathbb{R}^{D\times Q}$ and the vectors $\beta^*\in \mathbb{R}^D$, $\eta_0 \in \mathbb{R}^Q$. We define $\epsilon := (\epsilon_X,\epsilon_Y,\epsilon_H) \in \mathbb{R}^{D+Q+1}$ as a random variable with finite second moments, where its three components are pairwise independent random vectors (one could equivalently represent the latent variable effect implicitly with a SEM defined only for $(\Xb, Y)$ where $\epsilon_X$ and $\epsilon_Y$ are correlated). We can write this model in a matrix form:
\begin{equation}\label{equation-linear-sem}
    \begin{pmatrix}
    \Xb \\ Y \\ \Hb 
    \end{pmatrix} \!=\! \mathbf{B} \!\begin{pmatrix}
    \Xb \\ Y \\ \Hb 
    \end{pmatrix} \!+ \!\begin{pmatrix}
    \epsilon_X \\ \epsilon_Y \\ \epsilon_H 
    \end{pmatrix}\!; \; \mathbf{B}\! = \!\begin{pmatrix}
    B_{XX} & 0 & B_{XH}
    \\ \beta^{*T} & 0 & \eta_0^T
    \\ 0 & 0 & 0
    \end{pmatrix}
\end{equation}
where $(\Sigma_{\epsilon_X},\Sigma_{\epsilon_Y}, \Sigma_{\epsilon_H})$ are the covariance matrices of $\epsilon_X,\epsilon_Y,\epsilon_H$ respectively. We assume that $(I -\mathbf{B})$ is invertible, and thus the solution $\pi$ to the system is characterized by $\mathbf{B}$ and the distribution of $\epsilon$ through $\pi := \mathcal{D}\big((I -\mathbf{B})^{-1}\epsilon\big)$ where we denote by $\mathcal{D}(U)$ the distribution of the random variable $U$. The SEM model presented above can be reformulated with a probabilistic graphical model, where the variables (both observed or latent) correspond to the nodes and the edges are determined by the connectivity matrix $\mathbf{B}$. Our objective is to identify the parameter $\beta^*$ that captures the causal dependence between the observed covariates and the response, and especially to identify the non-zero entries: this is the problem of causal discovery. In the context of graphical models we want to identify the parent nodes of the response node, especially so when $\beta^*$ is assumed to be sparse and it is of importance to identify the few causal factors. Based only on $\Xb,Y$ samples from $\pi$, the least squares regression vector of $Y$ on the covariates $\Xb$ is equal to $\beta^*$ plus an additive confounding term (we provide an explicit formulation in Prop.~\ref{proposition:covariance-parameters}). Our method recovers $\beta^*$ by leveraging information from multiple environments, and under sparsity assumptions we propose a regularized version to specifically identify relevant covariates. Environments in practice correspond to different data sources: if the collected data are patient medical records, different hospitals may correspond to different environments. When analyzing cellular genetic pathways, measurements from different cell cultures where different genes have been knocked out correspond to data from different environments. The crucial \emph{invariance} assumption is that the mechanism generating the response variable $Y$, encoded in the parameters of the structural equation of $Y$, is constant across environments. In particular, those parameters relating the response to the observed covariates, encoded by $\beta^*$, are invariant. We observe data from $E\geq 1$ different environments, which are characterized by their underlying respective distributions over $\mathbb{R}^D \times \mathbb{R}$. We use the subscript $(\Xb_e, Y_e)$ to refer to the observed variables in environment $e \in [E]:=\{1,\dots,E\}$ (and no longer to the covariate index as in \ref{equation-linear-sem-individual}). Our main assumption defines how models relate to each other -- i.e. which elements in equation~\ref{equation-linear-sem} may change--, which we call a \emph{model of environments}. This represents our beliefs on how interventions may perturb the initial observational data generating distribution. In our work, we only allow for $(\Sigma_{e,\epsilon_X}, \Sigma_{e,\epsilon_Y})$ to depend on the environment, so that each $e$-th environment distribution $\pi_e$ is the solution of the linear SEM:
\begin{equation}\label{eq:model-of-environments}
    \begin{pmatrix}
    \Xb_e \\ Y_e \\ \Hb 
    \end{pmatrix} = \mathbf{B} \begin{pmatrix}
    \Xb_e \\ Y_e \\ \Hb 
    \end{pmatrix} + \begin{pmatrix}
    \epsilon_{e,X} \\ \epsilon_{e,Y} \\ \epsilon_H 
    \end{pmatrix}
\end{equation}
This model of environments can be understood as transformations from a base SEM via ``shift interventions'' on variables $\Xb,Y$ \citep{rothenhausler2015backshift, rothenhausler2019causal}, where an environment-specific additive noise term is added to $\epsilon$ in the SEM: the distribution of the $e$-th environment is given by $\pi_e := \mathcal{D}\big((I -\mathbf{B})^{-1}(\epsilon + v_e)\big)$ where $v_e = (v_{e,X}, v_{e,Y}, 0)$ is a centered random variable in $\mathbb{R}^{D+Q+1}$ where only the first $D + 1$ coordinates are non-zero, independent of $\epsilon$. These interventions are also called ``parametric'' interventions (on $\Xb$ and $Y$) \citep{eberhardt2007interventions} or ``soft'' interventions \citep{eaton2007exact}. Constraining $v_{e,Y} = 0$ would correspond to shift interventions only on $\Xb$ (cf. next paragraph and ``Shift $\Xb$'', ``Shift $Y$'' in Table~\ref{table-methods} for further details). These contrast with ``stuctural'' or ``perfect'' interventions that also allow for modifications in the matrix $\mathbf{B}$, which occur for example with do-interventions \citep{judea2000causality}. A graphical representation of this model of environments is presented in Figure~\ref{fig:DAG}: different environments are obtained via changes in the distributions of $\epsilon_X, \epsilon_Y$ via shift interventions, represented via red circles. The remaining structure of the model is unchanged across environments. Defining a model of environments carries a trade-off: the smaller the set of elements that change across environments is, the easier it is to find invariant quantities across environments, but conversely that corresponds to stronger assumptions that will restrict the scope of applications. From now on we no longer explicitly mention the latent variables and hence $\pi_e$ refers to the marginal distribution over $\Xb_e, Y_e$.
\begin{table}
  \caption{Multi-environment analysis methods and their scope of application.}
  \label{table-methods}
  \centering
  \begin{tabular}{llllll}
    Method     & Shift $\Xb$ & Shift $Y$ & Confounding & Recovers $\beta^*$\\
    \midrule
    KL Regression  & Yes & Yes & Yes & Yes \\
    Invariant Causal Prediction  & Yes & No &  No & Yes \\  
    Causal Dantzig  & Yes & Yes & Yes & Yes\\
    Invariant Risk Minimization  & Yes & Yes & No & No \\
    Anchor Regression  & Yes & Yes & Yes & No\\
    Causal Transfer Learning & Yes & No & No & Yes \\
    \bottomrule
  \end{tabular}
\end{table}
\paragraph{Related Literature}
The idea that true causal mechanisms are \emph{invariant} across different experimental settings \citep{peters2016causal, buhlmann2018invariance} has generated a lot of interest recently in the causal inference community. Based on samples collected from different ``environments'', it is possible to identify covariates whose effect on the response variable is universal, which are defined as causal. Solving the causal discovery problem problem with purely observational data is not possible, and a popular line of work in causal inference which uses Bayesian networks to represent causal models \citep{pearl1995causal,spirtes2000causation} relies on experimental or interventional data recover the graph structure. Pearl's do-calculus \citep{judea2000causality} identifies causal effects based on observational data but assuming perfect knowledge of the graphical structure of the Bayesian network. Other methods assume different data sources are generated through a change in the structure of the Bayesian network, and under some assumptions the causal graph can be recovered \citep{eberhardt2010combining,hyttinen2012learning}. This requires detailed knowledge on the perturbations that originate the distinct data sources. On the other hand, under a more limited knowledge of the mechanism that generates different data sources, the different distributions can be modeled with ``uncertain'' interventions \citep{korb2004varieties,eaton2007exact,eberhardt2010combining,tian2013causal}, for which we do not have a clear knowledge, and is the setting we assume in this work. Also, instead of estimating the complete graph structure \citep{spirtes2000causation,chickering2002optimal,hauser2012characterization,rothenhausler2015backshift}, we rather focus on identifying the sub-graph that connects the response variable to the other covariates. This leads us to formulating a linear SEM as in equation~\ref{equation-linear-sem}, and we now present some methods that solve this problem (we provide in Table~\ref{table-methods} a quick summary of how these relate to each other). The main modeling assumption for any procedure leveraging multiple environments is the definition of how such environments relate one to another, i.e. what the \emph{model of environments} is. Usually, procedures that focus on recovering $\beta^*$ (using our linear SEM framework) usually do not allow for any modification of the structural equation of $Y$: this is the basis of Invariant Causal Prediction (ICP) \citep{peters2016causal} and further methods based on the same principle \citep{heinze2018invariant,pfister2019invariant}. This body of work is based on the assumption that the conditional distribution of the response on a subset of covariates that are termed ``causal'' is constant across environments: i.e. if $S^*\subset[D]$ are the set of indices of such causal variables, denoting $\Ub_{A}:= (U_a)_{a\in A}$ for a vector $\Ub$, and $A$ subset of indices, then 
\begin{equation}\label{invariant-assumption}
    Y_e|\Xb_{e,S^*} \stackrel{d}{=}Y_f|\Xb_{f,S^*} \quad \forall e, f\in [E]
\end{equation}{}
This assumption, translated into our SEM model, implies that the row corresponding to $Y$ is invariant across environments (i.e. no ``shift $Y$'' interventions in Table~\ref{table-methods}), hence our assumptions are weaker in that aspect. This weakened assumption for the model of environments is in line with what methods such as Invariant Risk Minimization (IRM) \citep[Definition 6, p.10]{arjovsky2019invariant} assume and argue is needed in real-world applications: changes in the noise level of the response across environments. The structure relating the covariates to the response---within the connectivity matrix $\mathbf{B}$---is invariant in any case (indeed, $\beta^*$ is defined because of such invariance). However, our method fundamentally differs from those other methods that allow shifts in $Y$, as their goal is not to provide an estimate of $\beta^*$ (i.e. ``Recovers $\beta^*$'' in Table~\ref{table-methods}), but instead solve distributional robustness problems:
\begin{align*}
    \hat{\beta} := \min_{b \in \mathbb{R}^D} \max_{e\in [E]} \mathbb{E}[(Y_e - \Xb_e b)^2]
\end{align*}{}
These estimators minimize the worst-case prediction risk, and are related to adversarial training \citep{sinha2017certifying,gao2017wasserstein}. Whenever the set of environments contains all possible modifications of the structural equations except that of the target, the output of such optimization problem is $\beta^*$, as described in Causal Transfer Learning (CTL) \citep{rojas2018invariant}. Besides this setting, theoretical guarantees of those estimators will usually focus on hedging against a particular subset of interventions instead of estimating $\beta^*$, as in Anchor Regression \citep{rothenhausler2018anchor}. Finally, going back to those methods that focus on recovering $\beta^*$, most of them do not allow for confounding (i.e. ``Confounding'' in Table~\ref{table-methods}). The invariance assumption~(\ref{invariant-assumption}) fails whenever a latent variable $\Hb$ simultaneously affects $Y$ or $\Xb$, or whenever its distribution changes across environments \citep[Example 1, p.6]{christiansen2020switching}. Several of the previously mentioned techniques fail to recover $\beta^*$ under confounding, and recently proposed methods such as \cite{christiansen2020switching} restrict the latent variables to be discrete (and do not allow for changes in the noise level $\Sigma_{\epsilon_Y}$ of the response). KL regression shares many characteristics with Causal Dantzig \citep{rothenhausler2019causal}, which has an unregularized and a sparsity-inducing regularization versions. These two methods are related in a way that is analogous to the relationship between the Lasso linear regression \citep{tibshirani1996regression} and the Dantzig Selector \citep{candes2007dantzig}. Causal Dantzig without regularization leads to an estimator that is similar to ours in its scope of applications. However, as we show in the experiments, KL regression outperforms unregularized Causal Dantzig in terms of stability. Using a sparsity-inducing regularization in Causal Dantzig greatly improves its behavior and matches KL regression performance. However, this regularized Causal Dantzig \emph{needs} sparsity in $\beta^*$, otherwise it does not properly recover it. KL regression does not require sparsity to perform well, although the regularized KL regression version substantially outperforms any other method in sparse settings.

\subsection{Invariant Elements for our Model of Environments}

Assume that variables $\epsilon$ defined in the SEM (cf. equation~\ref{equation-linear-sem}) have finite second moments. Denote by $\Sigma_e^{joint}$ the covariance matrix of the observed variables in environment $e$, which we assume has full rank.
\begin{align*}
    \Sigma_e^{joint} := \begin{pmatrix}
    \Sigma_{e,X} & \Sigma_{e,XY}
    \\ \Sigma_{e,XY}^T & \Sigma_{e,Y}
    \end{pmatrix} := \text{Cov}\begin{pmatrix}
    \Xb_e
    \\Y_e
    \end{pmatrix} 
\end{align*}
Our model of environments assumption, that relies on a linear SEM, leads to a specific structure of these covariances across environments. Our method recovers $\beta^*$ only via the analysis of $\big(\Sigma_e^{joint}\big)_{e \in [E]}$, i.e. the second moments, regardless of the specific distribution of $\epsilon$. As a first step, we observe that in any given environment we can decouple the confounding bias from $\beta^*$ in the regression vector estimated by least squares.
\begin{proposition}\label{proposition:covariance-parameters}
Under our model of environments (cf. equation~\ref{eq:model-of-environments}) we have:
\begin{align*}
    & \Sigma_{e,X} =  C\Big(\Sigma_{e,\epsilon_{X}} + B_{XH}\Sigma_{\epsilon_H}B_{XH}^T\Big)C^T 
    \\ & \Sigma_{e,XY} = \Sigma_{e,X}\beta^* + \eta^*
\end{align*}{}
where $C := (I -B_{XX})^{-1}$ and $\eta^* := CB_{XH}\Sigma_{\epsilon_{H}}\eta_0$.
\end{proposition}
We prove this result in Appendix~\ref{proof:proposition:covariance-parameters}. In particular, $\Sigma_{e,X}$ depends on the environment $e$ through $\Sigma_{e, \epsilon_X}$, and $\beta^*, \eta^*$ do not depend on $e$. Given data from the $e$-th environment, the least squares estimator $\hat{\beta}_e$ estimates the population regression coefficient given by $\beta_e := \Sigma_{e,X}^{-1}\Sigma_{e,XY} = \beta^* + \Sigma_{e,X}^{-1}\eta^*$. The confounding due to the additional term $\Sigma_{e,X}^{-1}\eta^*$ implies that, based on purely observational data (i.e. data from one single environment), least squares is biased for $\beta^*$.

\section{Kullback-Leibler Regression}\label{section:section-3}

\subsection{General Framework}
We reformulate the joint covariance of environment $e$ through the regression coefficient $\beta_e$ and the residual variance given by \mbox{$\sigma_e^2 := \Sigma_{e,Y} - \beta_e^T\Sigma_{e,X}\beta_e \in \mathbb{R}$}: 
\begin{align*}
    \Sigma_e^{joint} = 
    \begin{pmatrix}
    \Sigma_{e,X} & \Sigma_{e,X}\beta_e
    \\ \beta_e^T\Sigma_{e,X} & \sigma_e^2 + \beta_e^T\Sigma_{e,X}\beta_e
    \end{pmatrix}
\end{align*}
Let $\Pi$ be a mapping that we call the \emph{regression function} associated to the $e$-th environment. Its input consists on the pair $(\Sigma_{e,X}, \sigma_e^2)$ --that we call \emph{regressors}, and the \emph{regression coefficient} $\theta$:
\begin{equation}\label{distribution-map}
    \Pi:\begin{cases}{}\big(\mathbb{R}^{D\times D}\times\mathbb{R}_+\big)\times\mathbb{R}^D \rightarrow \mathbb{R}^{(D+1)\times (D+1)} 
    \\(\Sigma_{e,X}, \sigma_e^2), \theta \mapsto
    \begin{pmatrix}
    \Sigma_{e,X} & \Sigma_{e,X}\theta
    \\ \theta^T\Sigma_{e,X} & \sigma_e^2 + \theta^T\Sigma_{e,X}\theta
    \end{pmatrix}
    \end{cases}
\end{equation}{}
The output is a positive semi-definite matrix such that $\Sigma_e^{joint} = \Pi\big((\Sigma_{e,X}, \sigma_e^2),\beta_e\big)$. For notation simplicity, let $\Pi_e(\theta):=\Pi\big((\Sigma_{e,X}, \sigma_e^2),\theta\big)$. Given a distance $\delta(\cdot, \cdot)$ over the set of positive semi-definite matrices, we can construct a loss function over $\theta$:
\begin{align}\label{loss-eq-simple}
    & \mathcal{L}(\theta) := \sum_{e\in [E]} \delta\big(\Pi_e(\theta), \Sigma_e^{joint}\big)
\end{align}{}
Although the distributions $\pi_e$ are not Gaussian, we can follow a quasi-maximum likelihood approach and define $\delta$ as the Kullback-Leibler divergence between two multivariate Gaussian variables with covariances $\Pi_e(\theta)$ and $\Sigma_e^{joint}$. This leads to the following environment-wise Mahalanobis distance in $\theta$ parametrized by the regressors $\Sigma_{e,X},\sigma_e^2$.
\begin{align*}
    \delta\big(\Pi_e(\theta), \Sigma_e^{joint}\big) := KL\Big(\mathcal{N}\big(\Pi_e(\theta)\big)\Big\Vert \mathcal{N}\big(\Sigma_e^{joint}\big)\Big)
\end{align*}
We provide in Appendix~\ref{appendix:regression-analogy} additional insights on this choice of distance and why our method can be understood as performing linear regression in a space of distributions. We then construct an estimator $\hat{\theta}$ by minimizing the corresponding empirical risk by plugging-in the empirical counterparts of $\Sigma_{e,X}, \sigma_e^2,\beta_e$ (denoted by a hat):
\begin{align}\label{loss-eq-simple-empirical}
    & \hat{\mathcal{L}}(\theta) := \sum_{e\in [E]} \delta\Big(\Pi\big((\hat{\Sigma}_{e,X},\hat{\sigma}_e^2), \theta\big), \hat{\Sigma}_e^{joint}\Big)
    \\ & \hat{\theta} := \argmin_{\theta\in \mathbb{R}^D} \hat{\mathcal{L}}(\theta)
\end{align}{}
This particular optimization problem in $\theta$ is convex, and has a closed-form solution. This procedure looks for a parameter $\theta$ that makes the environment-wise distributions as compatible as possible with a \emph{model of distributions}. Here, by minimizing $\mathcal{L}(\theta)$, we implicitly assume that there is a coefficient $\theta^*$ that defines for each environment $e$ a distribution $\pi_e^*$ whose covariance matrix is given by $\Pi_e(\theta^*)$. The observed environment distributions $\pi_e$ are then just ``unstructured'' perturbations of such $\pi_e^*$, that is there is no relationship between the regressors (i.e. $(\Sigma_{e,X}, \sigma_e^2)$) and the perturbation of the coefficient $\beta_e$ with respect to $\theta^*$.  We here understand the role of the structural causal model: individually, each of the environment distributions do not allow us to recover the invariant vector. It is only through the joint analysis of the environments that we identify it, and thus one needs to account for this causal structure in the parametrization in the loss. The problem formulation above is not adapted to our setting where the confounding bias in $\beta_e$ is environment-dependent, related to the regressors.

\subsection{Kullback-Leibler Regression for our Model of Environments}

\begin{figure*}[t]
\centering
\includegraphics[width=0.9\linewidth]{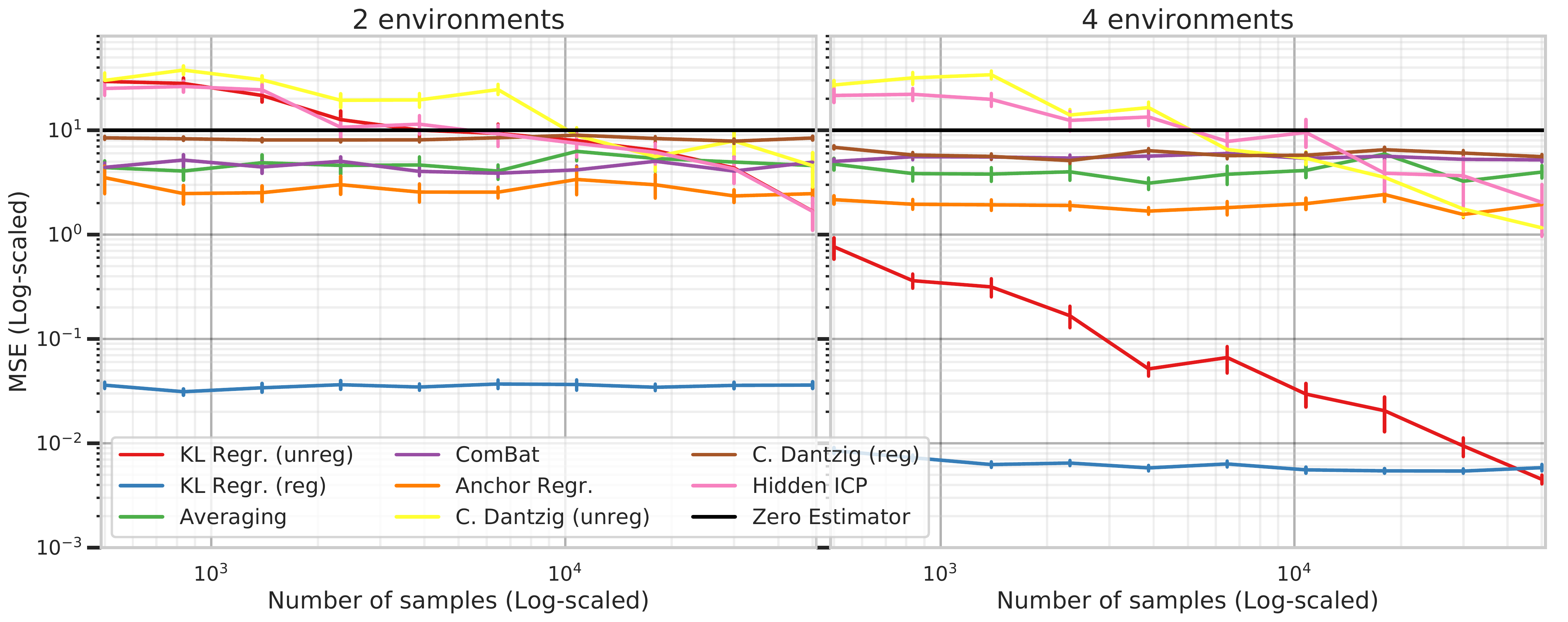}
\caption{\textbf{Baseline comparison} Plots of the (log-scaled) MSE of the estimator of the regression parameter (with respect to the true $\beta^*$) as a function of the number of samples. The $L_1$ regularized KL regression consistently achieved the lowest MSE.}
\label{fig:figure_a}
\end{figure*}

CV the choice of lambda for KL regression to get also a decrease with number of samples.

Whenever our environment distributions $\pi_e$ come from a latent variable model as in the previous section (cf. equation~\ref{eq:model-of-environments}), minimizing the parameter $\theta$ in the mapping $\Pi_e(\theta)$ in \eqref{distribution-map} that defines our loss function does not yield the desired result, basically because there is no $\theta^*$ that is equal to all $\beta_e$ simultaneously. However, as we saw in Proposition~\ref{proposition:covariance-parameters}, the expression of the confounded linear regression coefficient $\beta_e$ in environment $e$ can be decomposed into two terms $\beta_e = \beta^* + \Sigma_{e,X}^{-1}\eta^*$, where $\eta^*$ no longer depends on $e$. As such, we can modify the loss in \eqref{loss-eq-simple} by expressing $\theta = \beta + \Sigma_{e,X}^{-1}\eta$, and then minimize the reparametrized loss over $(\beta,\eta)$, which is still jointly convex in both arguments:
\begin{align}\label{loss-DR}
    & \mathcal{L}(\beta, \eta) \!:=\! \sum_{e\in [E]}\! \delta\Big(\!\Pi_e(\beta + \Sigma_{e,X}^{-1}\eta))\Vert \Sigma_e^{joint}\!\Big)
    \\ & (\tilde{\beta}^{KL}, \tilde{\eta}) := \argmin_{\beta, \eta \in \mathbb{R}^D} \mathcal{L}(\beta, \eta)
\end{align}{}
Importantly, the reparametrization only requires knowledge of the covariance of $\Xb_e$, which can be estimated. This is the key insight of our method: because of the invariant structure of the model of environments and the judicious choice of distance in the space of covariances, we can construct a tractable loss that is minimized at the invariant vector. We denote by $\tilde{\beta}^{KL}$ the solution to the above optimization problem, and look for conditions that guarantee uniqueness of the solution, thus identifying $\beta^*$.
\begin{proposition}\label{proposition:solution-DR}
Provided that the following matrix is invertible
\begin{align}
    S_{\beta}:= \Big(\sum_{e\in [E]}\frac{\Sigma_{e,X}^{-1}}{\sigma_e^2}\Big)\Big(\sum_{e\in [E]}\frac{\Sigma_{e,X}}{\sigma_e^2}\Big) - \Big(\sum_{e\in [E]}\frac{1}{\sigma_e^2}\Big)^2I_D
\end{align}
the unique solution to the KL regression problem~\ref{loss-DR} has a closed form given by 
\begin{align*}
    \tilde{\beta}^{KL} = S_{\beta}^{-1}\Bigg(&\Big(\sum_{e\in [E]}\frac{\Sigma_{e,X}^{-1}}{\sigma_e^2}\Big)\Big(\sum_{e\in [E]}\frac{\Sigma_{e,X}\beta_e}{\sigma_e^2}\Big)  -  \Big(\sum_{e\in [E]}\frac{1}{\sigma_e^2}\Big)\Big(\sum_{e\in [E]}\frac{\beta_e}{\sigma_e^2}\Big)\Bigg)
\end{align*}{}
Assuming that our environment distributions $\pi_e$ follow the SEM model above, then we have $\tilde{\beta}^{KL} = \beta^*$ (at the population level, i.e. assuming we can perfectly estimate the joint covariance under $\pi_e$).
\end{proposition}{}
\begin{proof}\label{proof:proposition:distribution-DR}
We prove here Proposition~\ref{proposition:distribution-DR}. Again, we drop the subscript $X$ from the covariance matrix $\Sigma_{e,X}$ for notation simplicity. Let the following matrices in $\mathbb{R}^{dE\times d}$:
\begin{align*}
    \mathbb{U} =  \begin{pmatrix}{}
    \frac{\hat{\Sigma}_1}{\sigma_1}
    \\\vdots 
    \\ \frac{\hat{\Sigma}_E}{\sigma_E}
    \end{pmatrix}; \quad 
    \mathbb{V} =  \begin{pmatrix}{}
    \frac{\hat{\Sigma}^{-1}_1}{\sigma_1}
    \\\vdots 
    \\ \frac{\hat{\Sigma}^{-1}_E}{\sigma_E}
    \end{pmatrix} ; \quad 
    \mathbb{I} =  \begin{pmatrix}{}
    \frac{I_D}{\sigma_1}
    \\\vdots 
    \\ \frac{I_D}{\sigma_E}
    \end{pmatrix}
\end{align*}{}
We have that 
\begin{align*}
    \hat{S}_{\beta} = & \mathbb{I}^T\mathbb{V}\mathbb{U}^T\mathbb{I} - \mathbb{I}^T\mathbb{I}\mathbb{I}^T\mathbb{I}
    \\ \hat{\beta} = & (\mathbb{I}^T\mathbb{V}\mathbb{U}^T\mathbb{I} - \mathbb{I}^T\mathbb{I}\mathbb{I}^T\mathbb{I})^{-1}(\mathbb{I}^T\mathbb{V}\mathbb{U}^T - \mathbb{I}^T\mathbb{I}\mathbb{I}^T)\hat{\mathbb{B}}
    \\ = & \mathbb{W}\hat{\mathbb{B}}
\end{align*}{}
where 
\begin{align*}
    \mathbb{W} = & (\mathbb{I}^T\mathbb{V}\mathbb{U}^T\mathbb{I} - \mathbb{I}^T\mathbb{I}\mathbb{I}^T\mathbb{I})^{-1}(\mathbb{I}^T\mathbb{V}\mathbb{U}^T - \mathbb{I}^T\mathbb{I}\mathbb{I}^T) \in \mathbb{R}^{d\times dE}
    \\ \hat{\mathbb{B}} = & \begin{pmatrix}{}
    \frac{\hat{\beta}_1}{\sigma_1}
    \\ \vdots
    \\ \frac{\hat{\beta}_E}{\sigma_E}
    \end{pmatrix} \in \mathbb{R}^{dE}; \qquad
    \mathbb{B} =  \begin{pmatrix}{}
    \frac{\beta_1}{\sigma_1}
    \\ \vdots
    \\ \frac{\beta_E}{\sigma_E}
    \end{pmatrix} \in \mathbb{R}^{dE}
\end{align*}
so that each stacked vector in $\hat{\mathbb{B}}$ is independent of the others, and for every $e\in [E]$, we have that $\hat{\beta}_e = \beta_e + \frac{\sigma_e}{\sqrt{n_e}}\hat{\Sigma}^{-1/2}_{e,X}z_e$ where $(z_e)_{1\leq e \leq E}$ are independent standard Gaussians. We therefore get that 
\begin{align*}
    \hat{\mathbb{B}} = & \begin{pmatrix}{}
    \frac{\beta_1}{\sigma_1}
    \\ \vdots
    \\ \frac{\beta_E}{\sigma_E}
    \end{pmatrix} + \begin{pmatrix}
    \frac{1}{\sqrt{n_1}}\hat{\Sigma}_1^{-1/2}z_1
    \\ \vdots
    \\ \frac{1}{\sqrt{n_E}}\hat{\Sigma}_E^{-1/2}z_E
    \end{pmatrix} = \mathbb{B} + \mathbb{Z}
\end{align*}{}
Therefore $\hat{\beta}^{KL} = \mathbb{W}\hat{\mathbb{B}} = \mathbb{W}\mathbb{B} + \mathbb{W}\mathbb{Z}$. But we have that $\mathbb{W}\mathbb{B} = \mathbb{E}[\hat{\beta}^{KL}|(X_e)_e]$, and $\mathbb{W}\mathbb{Z}$ is a multivariate Gaussian, whose covariance is given by :
\begin{align*}
    V = & \mathbb{W}\begin{pmatrix}{}
    \frac{1}{n_1}\hat{\Sigma}^{-1}_1 & 0&0 &\cdots & 0
    \\0 & \frac{1}{n_2}\hat{\Sigma}^{-1}_2& 0 &\cdots & 0
    \\ 0 & 0 & \ddots& \cdots & 0
    \\ \vdots & \vdots&&\vdots&
    \\ 0 & \cdots & 0& 0 &\frac{1}{n_E}\hat{\Sigma}_E^{-1}
    \end{pmatrix}\mathbb{W}^T
    \\ = & \sum_{e=1}^E \frac{1}{n_e}A_e
\end{align*}{}
for some symmetric matrix $A_e$. Now, if all the environments have the same sample size, $n_e = n$, then
\begin{align*}
    V & = \frac{1}{n}\hat{S}_{\beta}^{-1}\mathbb{I}^TA\mathbb{I}\hat{S}_{\beta}^{-1,T}
    \\ A & = (\mathbb{V}\mathbb{U}^T - \mathbb{I}\mathbb{I}^T)\begin{pmatrix}{}
    \hat{\Sigma}^{-1}_1 & 0&0 &\cdots & 0
    \\0 & \hat{\Sigma}^{-1}_2& 0 &\cdots & 0
    \\ 0 & 0 & \ddots& \cdots & 0
    \\ \vdots & \vdots&&\vdots&
    \\ 0 & \cdots & 0& 0 &\hat{\Sigma}_E^{-1}
    \end{pmatrix}(\mathbb{U}\mathbb{V}^T - \mathbb{I}\mathbb{I}^T)
    \\ & = \Big[\frac{\hat{\Sigma}_i^{-1}\hat{\Sigma}_j - I}{\sigma_i \sigma_j} \Big]_{i,j}\Big[ \hat{\Sigma}_j^{-1}1_{j=k}\Big]_{j,k}\Big[\frac{\hat{\Sigma}_k\hat{\Sigma}_l^{-1} - I}{\sigma_k \sigma_l} \Big]_{k,l}
    \\ & = \Big[ \sum_{j=1}^E \frac{\hat{\Sigma}_i^{-1}\hat{\Sigma}_j\hat{\Sigma}_l^{-1} - \hat{\Sigma}_i^{-1} - \hat{\Sigma}_l^{-1} + \hat{\Sigma}_j^{-1}}{\sigma_i\sigma_j^2\sigma_l}\Big]_{i,l}
\end{align*}
\begin{align*}
    \mathbb{I}^TA\mathbb{I} & = \sum_{i,j,l = 1}^E \frac{\hat{\Sigma}_i^{-1}\hat{\Sigma}_j\hat{\Sigma}_l^{-1} - \hat{\Sigma}_i^{-1} - \hat{\Sigma}_l^{-1} + \hat{\Sigma}_j^{-1}}{\sigma_i^2\sigma_j^2\sigma_l^2}
    \\  &=\Big(\sum_{e\in [E]}\frac{\hat{\Sigma}_e^{-1}}{\sigma_e^2}\Big)\Big(\sum_{f\in [E]}\frac{\hat{\Sigma}_f}{\sigma_f^2}\Big)\Big(\sum_{e\in [E]}\frac{\hat{\Sigma}_e^{-1}}{\sigma_e^2}\Big) - \Big(\sum_{e\in [E]}\frac{1}{\sigma_e^2}\Big)^2\Big(\sum_{e\in [E]}\frac{\hat{\Sigma}_e^{-1}}{\sigma_e^2}\Big) 
    \\  &=\Big(\sum_{e\in [E]}\frac{\hat{\Sigma}_e^{-1}}{\sigma_e^2}\Big)\hat{S}_{\beta}^{T}
\end{align*}{}
Therefore we conclude 
\begin{equation*}
V = \frac{1}{n} \hat{S}_{\beta}^{-1}\Big(\sum_{e\in [E]}\frac{\hat{\Sigma}_e^{-1}}{\sigma_e^2}\Big)
\end{equation*}
\end{proof}{}

The conditions for this proposition are mild: instead of assuming specific types of interventions for each individual covariate to guarantee identifiability of $\beta^*$ \citep{hyttinen2012learning, peters2016causal}, we just require diverse enough environments in terms of the covariates covariance. Identifiability of $\beta^*$ holds as long as $S_{\beta}$ is invertible, regardless of the actual distribution of $\epsilon$. Under which conditions is the matrix $S_{\beta}$ invertible? In particular, if only one environment is available then $S_{\beta}$ is not invertible: we can not make any statement based on purely observational data. However, KL regression can output $\tilde{\beta}^{KL}$ based on only two different environments under some conditions, where all covariates are simultaneously perturbed in one of the two environments. We now reformulate a result from \cite{bhagwat_subramanian_1978} under our framework, which we prove in Appendix~\ref{proof:proposition:invertibility_S} for completeness.
\begin{proposition}[\cite{bhagwat_subramanian_1978}]\label{proposition:invertibility_S}
A sufficient condition for $S_{\beta}$ to be invertible is that there exists $e^*\in [E]$ such that the following matrix is invertible
\begin{align*}
\Sigma_{e^*,X}^{-1}  - \Big(\sum_{e\in [E]}\frac{1}{\sigma_e^2}\Big)^{-1}\Big(\sum_{e\in [E]}\frac{\Sigma_{e,X}^{-1}}{\sigma_e^2}\Big)
\end{align*}
\end{proposition}{}
Assume for simplicity that the residual variances are equal to 1 (corresponding to independent measurement errors), if $\Sigma_{2,X}^{-1}\Sigma_{1,X} - I_D$ is invertible then $S_{\beta}$ is invertible: consider the following example where such scenario may arise. A scientist can experimentally down-regulate simultaneously all (measured) expressions of genes of interest, without interfering with unobserved latent genes. Covariances of gene expressions will be on different scales (eg. $\Sigma_{1,X} \succ \Sigma_{2,X}$), and the latent confounding effect will remain unaltered (assuming a linear gene regulatory network model). For these two environments (observational and interventional), the condition above can be easily satisfied. More generally, simulations show that whenever the environments are more heterogeneous, KL regression performs better. 
In practice, all the quantities that define the loss $\mathcal{L}$ are estimated from the empirical covariances $\hat{\Sigma}_e^{joint}$: define $\hat{\beta}^{KL}$ as the plug-in estimator of $\tilde{\beta}^{KL}$ where we replace the environment population covariances (and derived quantities) by their empirical counterparts, where analogously $\hat{S}_{\beta}$ is the empirical version of $S_{\beta}$. For each environment $e \in [E]$ we observe \mbox{$n_e\geq 1$} i.i.d. samples. We can provide a distributional result: assuming that we have access to the population residual variances $\sigma_e^2$, conditioned on the sample covariates (i.e. as in fixed design linear regression), the KL regression estimator $\hat{\beta}^{KL}$ satisfies the following result:
\begin{proposition}\label{proposition:distribution-DR}
We have that, for $\epsilon$ Gaussian, 
\begin{align*}
    \hat{\beta}^{KL}\mid(X_e)_e \sim \mathcal{N}\big(\mathbb{E}[\hat{\beta}^{KL}|(X_e)_e], V \big)
\end{align*}{}
where $V = \sum_{e=1}^E \frac{1}{n_e}A_e$, for some symmetric matrices $A_e$ that depend only on the empirical covariance matrices. In the case where $n_e = n$ for all $e$, we have
\begin{align*}
V = \frac{1}{n} \hat{S}_{\beta}^{-1}\Big(\sum_{e\in [E]}\frac{\hat{\Sigma}_{e,X}^{-1}}{\sigma_e^2}\Big)
\end{align*}
\end{proposition}{}
We prove this proposition in Appendix~\ref{proof:proposition:distribution-DR}. If all environments have the same number of samples $n$, the plug-in estimator $\hat{\beta}^{KL}$ converges to $\tilde{\beta}^{KL}$ at rate $1/ \sqrt{n}$. Also, whenever $S_{\beta}$ is ill-conditioned, the variance of the estimator $\hat{\beta}^{KL}$ will be large. In general, we do not have access to the population residual variances $\sigma_e^2$, and the distribution is not Gaussian. We leave this extension to future work.

\subsection{Kullback-Leibler Regression with Sparsity Inducing Penalty}

\begin{figure*}[t]
\centering
\includegraphics[width=0.8\linewidth]{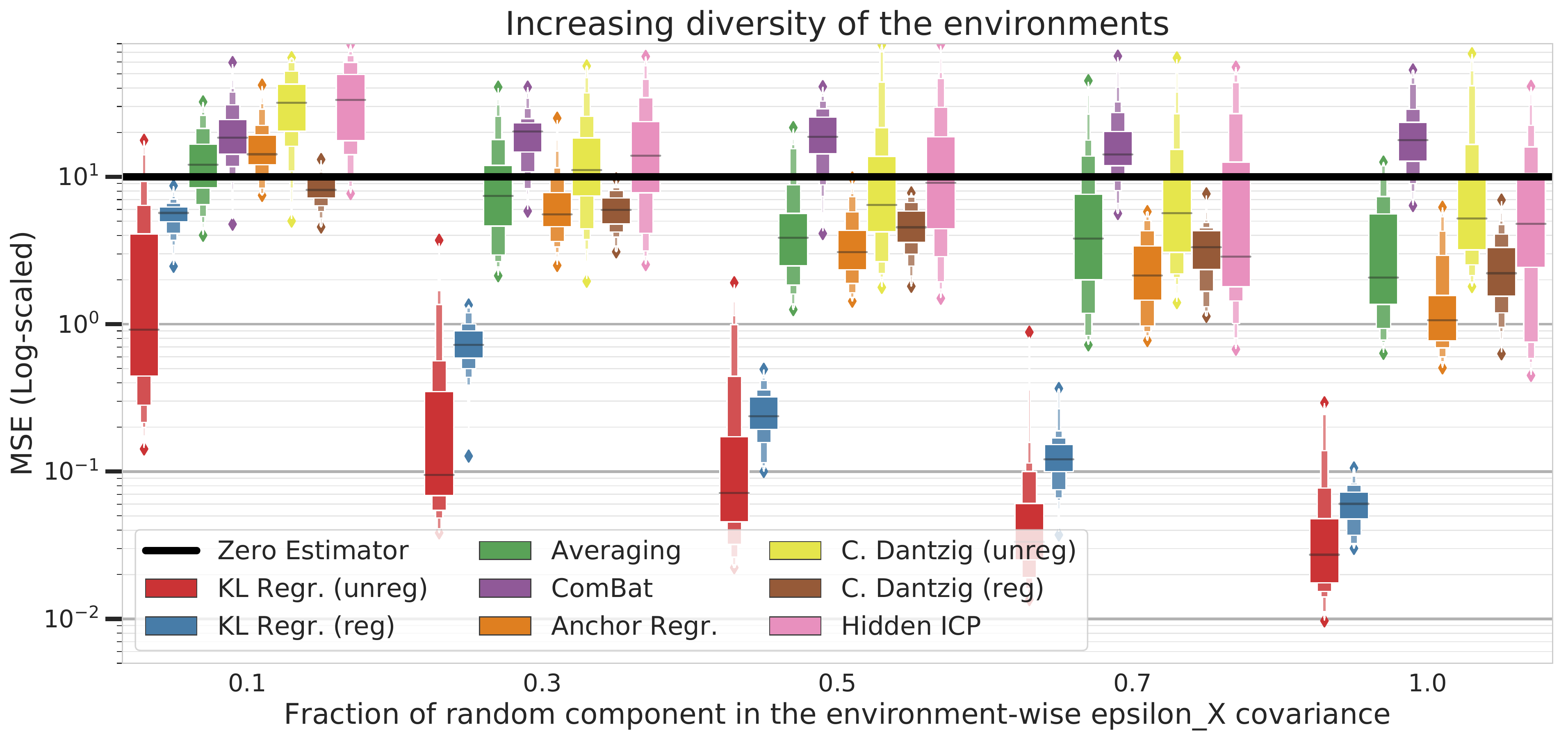}
\caption{\textbf{Diverse environments improve KL regression performance} Plots of the (log-scaled) MSE as a function of the environment-wise random component in $\Sigma_{e,\epsilon_X}$. }
\label{fig:environment-diversity}
\end{figure*}

Often times it is reasonable to assume that the invariant vector $\beta^*$ is sparse. Modifying KL regression to incorporate this prior knowledge can substantially improve the accuracy and stability of the method. We add an $L^1$-norm sparsity-inducing penalty to the optimization problem (\ref{loss-DR}), which is convex in $(\beta,\eta)$, to define the following \emph{Lasso KL regression}:
\begin{align}\label{loss-DR-sparse}
    & \mathcal{L}_{\lambda}(\beta, \eta) \!:=\!  \mathcal{L}(\beta, \eta) + \lambda\Vert\beta\Vert_1
    \\ & (\tilde{\beta}^{KL}(\lambda), \tilde{\eta}(\lambda)) := \argmin_{\beta, \eta \in \mathbb{R}^D} \mathcal{L}_{\lambda}(\beta, \eta)
\end{align}{}
Solving this problem in practice is easily done with off-shelf optimization packages such as CVXPY \citep{diamond2016cvxpy} in Python. Surprisingly, Lasso KL regression outperforms all the competing methods in sparse settings, is stable in small sample settings and robust to model misspecification. We leave for future work a deeper analysis of finite sample guarantees of the Lasso KL regression.

\subsection{Robustness of Kullback-Leibler Regression to Model Misspecification}

If we assume that the object of interest in the linear SEM (\ref{equation-linear-sem}), $\beta^*$, remains invariant across environments then we can bound its estimation error under misspecification. Assume we allow small perturbations in the remaining coefficients of our SEM model~(\ref{eq:model-of-environments}) for some of the environments: we define \mbox{$\eta_e := (I-B_{XX,e})^{-1}B_{XH,e}\Sigma_{\epsilon_X ,e}\eta_{0,e}$}. Define \mbox{$\delta_e := \eta_e - \eta^*$}, we have the following bound:
\begin{proposition}\label{proposition:robustness} We have:
\begin{align*}
& \Vert S_{\beta} (\tilde{\beta}^{KL} - \beta^*)\Vert^2 \leq C \sup_{e\in [E]}\Vert \delta_e\Vert^2
\end{align*}
where \mbox{$C:= \big(\sum_{e\in [e]}\frac{1}{\sigma_e^2}\big)\Big(\Vert\sum_{e\in [E]}\frac{\Sigma_e^{-1}}{\sigma_e^2}\Vert + \sum_{e\in [e]}\frac{\Vert\Sigma_e^{-1}\Vert}{\sigma_e^2}\Big)$}
\end{proposition}
In practice, we can bound the misspecification error in $\tilde{\beta}^{KL}$ with a bound on the (unknown) term $\sup_{e\in [E]}\Vert \delta_e\Vert^2$, as the remaining terms in $C$ can be estimated from the data. In particular, the effect of the model misspecification is entirely captured by $\delta_e$, even though the covariate covariance is also modified by such perturbations. 

\begin{figure}
     \centering
     \begin{subfigure}
\centering
\includegraphics[width=.8\linewidth]{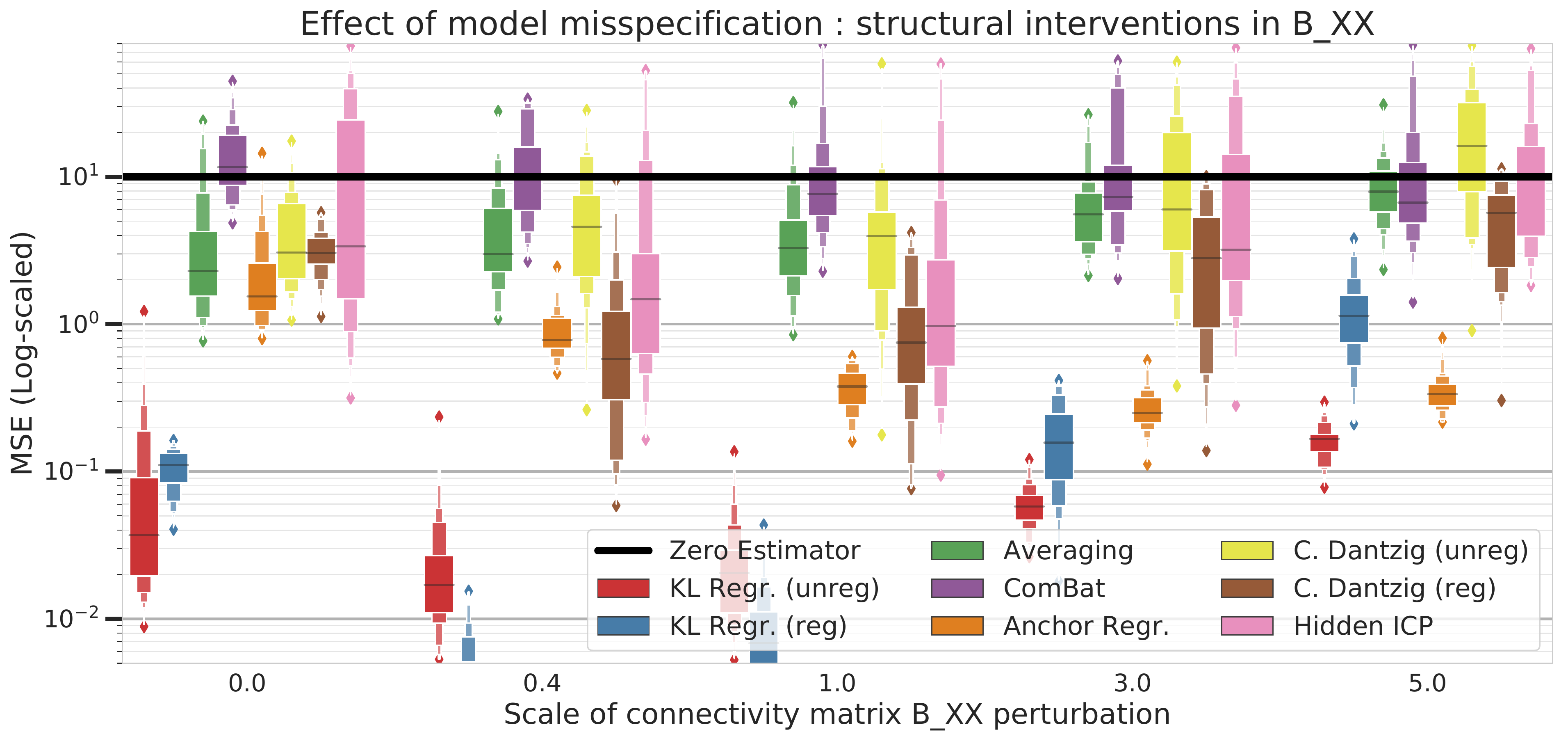}
\end{subfigure}
     \vfill
     \begin{subfigure}
\centering
\includegraphics[width=.8\linewidth]{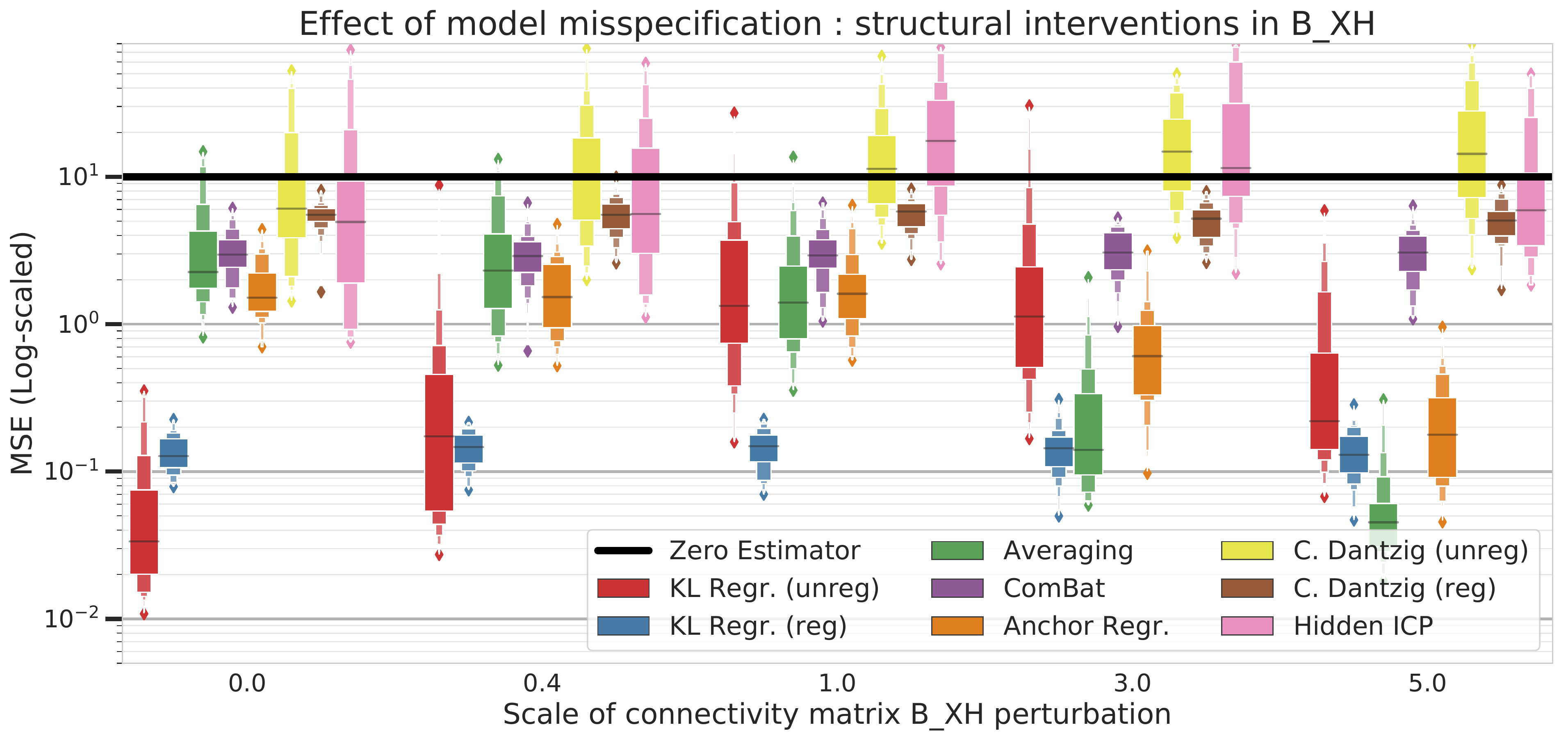}
\end{subfigure}
     \vfill
     \begin{subfigure}
\centering
\includegraphics[width=.8\linewidth]{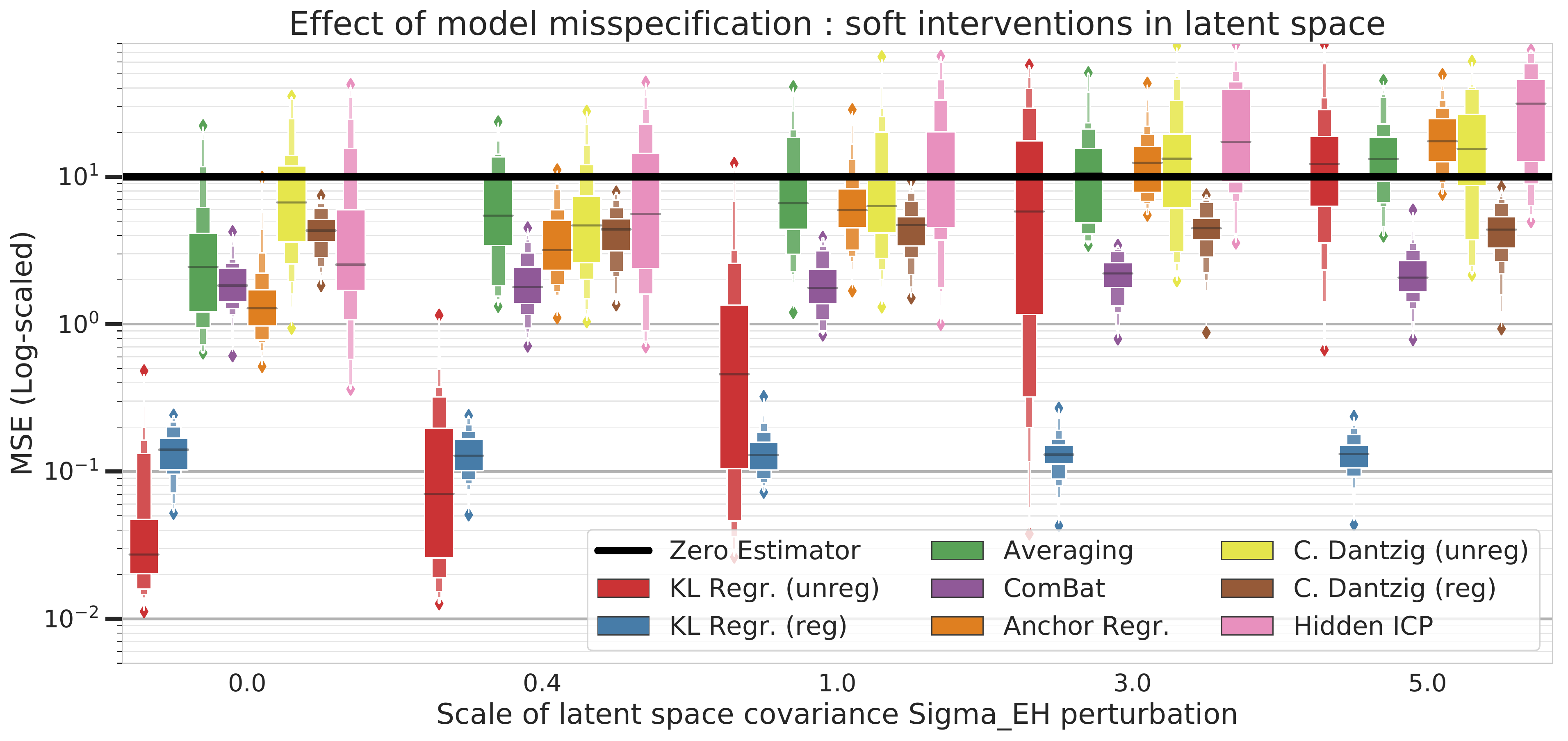}
\end{subfigure}
\label{fig:model-misspecification}
\caption{\textbf{Effect of model misspecification.} Plots of the (log-scaled) MSE of the estimator of the regression parameter (with respect to the true $\beta^*$) as a function of the noise level added to $B_{XX}$ (top), $B_{XH}$ (center) and $\Sigma_{\epsilon_H}$ (bottom). Surprisingly, KL regression performs remarkably well (in particular the $L^1$-penalized version), showing a certain level of robustness to model misspecification. The x-axis scaling is not linear.}
\end{figure}
Given that in practice this bound is hard to compute, we empirically evaluate the robustness of KL regression though simulations. We perturb the connectivity matrix $B_{XX}$ by adding a random noise to it, which varies in strength: denoting $s$ the scaling of the random component, we have that $B_{e,XX} = B_{0,XX} + sZ_e$, where $B_{0,XX}$ is shared across environments, and $Z_e$ is a matrix with standard Gaussian entries. We run analogous simulations perturbing the connectivity matrix $B_{XH}$, and perturbing the latent variable covariance matrix $\Sigma_{\epsilon_H}$. We report the results of the simulation in Figure~\ref{fig:model-misspecification}, where we vary $s \in [0,5]$ (where $s=0$ corresponds to our baseline configuration described in section~\ref{appendix:simulation-environments}) and select a few values that are representative of the general behavior of the estimators. As expected, KL regression is robust to small perturbations in the connectivity matrix and outperforms most other methods. As the strength of perturbation in $B_{e,XX}$ increases KL regression performance degrades substantially. Surprisingly, Lasso KL regression seems to be very robust to perturbations in $B_{XH}$, $\Sigma_{\epsilon_H}$. A noticeable effect is that, under a slight perturbation in $B_{XX}$, Anchor Regression improves substantially and performs surprisingly well, although it is designed only to protect against a certain shift-interventions in some particular directions. We hypothesize that the changes due to model misspecification lead to a more conservative hedging in multiple directions, bringing Anchor regression closer to Causal Transfer Learning \citep{rojas2018invariant}, which in theory recovers $\beta^*$.

\section{Experiments}\label{section:section-4}

\subsection{Validation on Synthetic Data}

We extensively evaluate KL regression on synthetic data. We compare the following methods: \textbf{1) KL regression}, both Lasso KL regression and unregularized KL regression. \textbf{2) Invariant Causal Prediction (ICP)} \citep{peters2016causal} The implementation we used in the R package (\texttt{InvariantCausalPrediction}) detects that the assumptions for using ICP are not satisfied, and suggests running the \texttt{hiddenICP} version. \textbf{3) Anchor Regression} \citep{rothenhausler2018anchor} \textbf{4) Causal Dantzig} \citep{rothenhausler2019causal}  We run experiments with both the regularized and unregularized versions. \textbf{5) ComBat} \citep{johnson2007adjusting} \textbf{6) Averaging the regression coefficients}. Appendix~\ref{appendix:simulation-environments} provides a detailed description of the data generating process of our synthetic data simulations, and precise description of all competing methods and the following simulations. 
\paragraph{Simulations with Varying Number of Environments.} Figure~\ref{fig:figure_a} plots the mean squared error (MSE) for the different estimators as a function of the number of samples for 2 and 4 environments. KL regression recovers $\beta^*$ with relatively few samples in several environments. Compared to alternatives in Table~\ref{table-methods}, KL regression is the only method that provably recovers $\beta^*$ at the population level. Compared to unregularized Causal Dantzig and hidden ICP (which behave similarly in 2 environments), KL regression is substantially more efficient as the number of environments increases. 
\paragraph{Increasing Diversity Across Environments.} As suggested in Proposition~\ref{proposition:invertibility_S}, having diverse covariance matrices $\Sigma_{e,X}$ implies that $S_{\beta}$ is likely to be invertible and well-conditioned, and thus the asymptotic variance of our KL regression estimator will be lower.  Figure~\ref{fig:environment-diversity} shows how MSE changes as the environment-wise covariances $\Sigma_{e,\epsilon_{X}}$ differ more one of another (cf. Appendix~\ref{appendix:diversity} for further details). All methods improve with the diversity of the environments, but KL regression particularly benefits from diversity. 

\paragraph{Additional Simulations.} We further explore the limitations and the robustness of our method in Appendix~\ref{appendix:simulations}, and summarize here our findings:
\begin{itemize}[noitemsep,nolistsep,leftmargin=*]
    \item We analyze the performance under various degrees of confounding  in Appendix~\ref{appendix:varying-confounding}. KL regression is robust as confounding gets stronger, while other methods degrade rapidly for higher confounding.
    \item We report in Appendix~\ref{appendix:sparsity} simulations where the sparsity level changes: regularized Causal Dantzig, and regularized KL regression which perfectly recover sparse vectors, degrade in settings where $\beta^*$ is not sparse. Unregularized KL regression works well in both sparse and non-sparse settings. 
    \item Artificially splitting the data from one environment to generate new environments does not help KL regression, it actually worsens it as it does not increase environment diversity (Appendix~\ref{appendix:split}).
    \item KL regression does not require data to originate from a Gaussian SEM. We validate the previous results on data sampled from Student's t-distribution with different degrees of freedom and report the results in Appendix~\ref{appendix:student}.
\end{itemize}

\subsection{Recovering Gene Regulatory Networks: DREAM Challenge Datasets}
Validating KL regression on real-world datasets is challenging as there is often no ground truth causal mechanism. Gene Regulatory Networks (GNR) can provide us with some partial ground truth, indicating only whether two genes, a transcription factor (TF) and a regulated gene, are related or not (i.e. whether the corresponding $\beta^*_i$ is non-null). We therefore focus on Lasso KL regression and regularized Causal Dantzig for this experiment, as they automatically select features. This type of datasets also fits well with our model of environments (cf. equation~\ref{eq:model-of-environments}): deletions on upstream genes provide the sort of perturbation that shifts the global covariate distribution of gene expressions, without altering the true regulation mechanism of a specific target gene. We run our method on some of the DREAM4 and DREAM5 datasets released as part of an well-known online challenge. Importantly, we can also benchmark KL Regression against the leaderboard of the challenge via the AUPR metric (area under precision-recall curve) used in the challenge. We were limited in our experiments to datasets for which the environment-wise covariance matrix $\Sigma_{e,X}$ is invertible, and we expect to relax this assumption in future work. We generated different environments depending on whether genes had been knocked-down. This required assumption on $\Sigma_{e,X}$ does not hold whenever the number of gene expression measurements is smaller than the number of TFs. However, single cell gene sequencing has the potential to make KL regression widely applicable in future GNR inference tasks. We provide in Appendix~\ref{appendix:DREAM} additional information for these real-data experiments.

In the well-known DREAM5 challenge, our method \emph{ranked at the top of the leaderboard} by a substantial margin for the analyzed dataset. For DREAM4, KL regression still ranked in the top 40\% of submissions: however the semi-synthetic datasets from DREAM4 contained one single knock-down experiments for every single gene, making our model of environments less realistic (compared to DREAM5 where a single knock-down experiment had multiple replicates).

\begin{table}
  \caption{AUPR Performance on DREAM Challenge.}
  \label{table:DREAM}
  \centering
  \begin{tabular}{lllllll}
    \toprule
    Method    & DREAM 4 & DREAM 5    \\
    \midrule
    Lasso KL Regr. & 0.26 & \textbf{0.45} \\
    C. Dantzig (reg.) &  0.15   & 0.27  \\
    Top Submission & \textbf{0.37} & 0.31 \\
    \bottomrule
  \end{tabular}
\end{table}

\section{Discussion}
We provide a detailed mathematical foundation for KL regression, as well as extensive validation on synthetic and real data. KL has strong mathematical guarantees, and works better across many settings compared to existing methods. We derived our theoretical guarantee for KL regression in a standard SEM setting commonly used in causal inference, without assuming that data is sampled from any particular parametric distribution (i.e. we do not assume data is Gaussian). Empirically, KL regression performs well even with large model mismatch; we are very encouraged that it \emph{achieved the top score on a Dream5 dataset}, which has complex nonlinear interactions. KL regression introduces a new regression-flavored perspective to the invariance-based causal inference framework, opening the door to potentially leverage other regression-related concepts such as bootstrapping, model selection techniques (AIC, BIC criterion) and various regularization methods (beyond the current Lasso KL regression $L^1$-penalized version). Interesting future research directions will also consist in developing a non-linear version of KL regression for generalized linear models. 

\subsubsection*{Acknowledgments}

The authors thank Dominik Rothenhäusler and Bryan He for their helpful comments, and Kailas Vodrahalli for his help with his implementation of invariant risk minimization. J.R.G. is supported by a Stanford Graduate Fellowship. J.Z. is supported by a Chan–Zuckerberg Biohub Investigator grant and National Science Foundation (NSF) Grant CRII 1657155.

\bibliography{main}

\begin{thebibliography}{42}
\providecommand{\natexlab}[1]{#1}
\providecommand{\url}[1]{\texttt{#1}}
\expandafter\ifx\csname urlstyle\endcsname\relax
  \providecommand{\doi}[1]{doi: #1}\else
  \providecommand{\doi}{doi: \begingroup \urlstyle{rm}\Url}\fi

\bibitem[Arjovsky et~al.(2019)Arjovsky, Bottou, Gulrajani, and
  Lopez-Paz]{arjovsky2019invariant}
Martin Arjovsky, L{\'e}on Bottou, Ishaan Gulrajani, and David Lopez-Paz.
\newblock Invariant risk minimization.
\newblock \emph{arXiv preprint arXiv:1907.02893}, 2019.

\bibitem[Ben-David et~al.(2010)Ben-David, Blitzer, Crammer, Kulesza, Pereira,
  and Vaughan]{ben2010theory}
Shai Ben-David, John Blitzer, Koby Crammer, Alex Kulesza, Fernando Pereira, and
  Jennifer~Wortman Vaughan.
\newblock A theory of learning from different domains.
\newblock \emph{Machine Learning}, 79\penalty0 (1-2):\penalty0 151--175, 2010.

\bibitem[Bhagwat and Subramanian(1978)]{bhagwat_subramanian_1978}
K.~V. Bhagwat and R.~Subramanian.
\newblock Inequalities between means of positive operators.
\newblock \emph{Mathematical Proceedings of the Cambridge Philosophical
  Society}, 83\penalty0 (3):\penalty0 393–401, 1978.

\bibitem[Bollen(1989)]{bollen1989structural}
Kenneth~A. Bollen.
\newblock \emph{Structural equations with latent variables}.
\newblock Wiley Series in Probability and Statistics, New Jersey, 1989.

\bibitem[B{\"u}hlmann(2018)]{buhlmann2018invariance}
Peter B{\"u}hlmann.
\newblock Invariance, causality and robustness.
\newblock \emph{arXiv preprint arXiv:1812.08233}, 2018.

\bibitem[Candes et~al.(2007)Candes, Tao, et~al.]{candes2007dantzig}
Emmanuel Candes, Terence Tao, et~al.
\newblock The dantzig selector: Statistical estimation when p is much larger
  than n.
\newblock \emph{The Annals of Statistics}, 35\penalty0 (6):\penalty0
  2313--2351, 2007.

\bibitem[Chickering(2002)]{chickering2002optimal}
David~Maxwell Chickering.
\newblock Optimal structure identification with greedy search.
\newblock \emph{The Journal of Machine Learning Research}, 3\penalty0
  (Nov):\penalty0 507--554, 2002.

\bibitem[Christiansen and Peters(2020)]{christiansen2020switching}
Rune Christiansen and Jonas Peters.
\newblock Switching regression models and causal inference in the presence of
  discrete latent variables.
\newblock \emph{Journal of Machine Learning Research}, 21\penalty0
  (41):\penalty0 1--46, 2020.
\newblock URL \url{http://jmlr.org/papers/v21/19-407.html}.

\bibitem[Diamond and Boyd(2016)]{diamond2016cvxpy}
Steven Diamond and Stephen Boyd.
\newblock Cvxpy: A python-embedded modeling language for convex optimization.
\newblock \emph{The Journal of Machine Learning Research}, 17\penalty0
  (1):\penalty0 2909--2913, 2016.

\bibitem[Eaton and Murphy(2007)]{eaton2007exact}
Daniel Eaton and Kevin Murphy.
\newblock Exact bayesian structure learning from uncertain interventions.
\newblock In \emph{Proceedings of the 11th International Conference on
  Artificial Intelligence and Statistics}, pages 107--114, 2007.

\bibitem[Eberhardt and Scheines(2007)]{eberhardt2007interventions}
Frederick Eberhardt and Richard Scheines.
\newblock Interventions and causal inference.
\newblock \emph{Philosophy of Science}, 74\penalty0 (5):\penalty0 981--995,
  2007.

\bibitem[Eberhardt et~al.(2010)Eberhardt, Hoyer, and
  Scheines]{eberhardt2010combining}
Frederick Eberhardt, Patrik~O. Hoyer, and Richard Scheines.
\newblock Combining experiments to discover linear cyclic models with latent
  variables.
\newblock In \emph{Proceedings of the Thirteenth International Conference on
  Artificial Intelligence and Statistics}, pages 185--192, 2010.

\bibitem[Fan et~al.(1999)Fan, Zhang, et~al.]{fan1999statistical}
Jianqing Fan, Wenyang Zhang, et~al.
\newblock Statistical estimation in varying coefficient models.
\newblock \emph{The Annals of Statistics}, 27\penalty0 (5):\penalty0
  1491--1518, 1999.

\bibitem[Gao et~al.(2017)Gao, Chen, and Kleywegt]{gao2017wasserstein}
Rui Gao, Xi~Chen, and Anton~J Kleywegt.
\newblock Wasserstein distributional robustness and regularization in
  statistical learning.
\newblock \emph{arXiv preprint arXiv:1712.06050}, 2017.

\bibitem[Gasperini et~al.(2019)Gasperini, Hill, McFaline-Figueroa, Martin, Kim,
  Zhang, Jackson, Leith, Schreiber, Noble, et~al.]{gasperini2019genome}
Molly Gasperini, Andrew~J Hill, Jos{\'e}~L McFaline-Figueroa, Beth Martin,
  Seungsoo Kim, Melissa~D Zhang, Dana Jackson, Anh Leith, Jacob Schreiber,
  William~S Noble, et~al.
\newblock A genome-wide framework for mapping gene regulation via cellular
  genetic screens.
\newblock \emph{Cell}, 176\penalty0 (1-2):\penalty0 377--390, 2019.

\bibitem[Hastie and Tibshirani(1993)]{hastie1993varying}
Trevor Hastie and Robert Tibshirani.
\newblock Varying-coefficient models.
\newblock \emph{Journal of the Royal Statistical Society: Series B (Statistical
  Methodology)}, 55\penalty0 (4):\penalty0 757--779, 1993.

\bibitem[Hauser and B{{\"u}}hlmann(2012)]{hauser2012characterization}
Alain Hauser and Peter B{{\"u}}hlmann.
\newblock Characterization and greedy learning of interventional markov
  equivalence classes of directed acyclic graphs.
\newblock \emph{Journal of Machine Learning Research}, 13\penalty0
  (79):\penalty0 2409--2464, 2012.

\bibitem[Heinze-Deml et~al.(2018)Heinze-Deml, Peters, and
  Meinshausen]{heinze2018invariant}
Christina Heinze-Deml, Jonas Peters, and Nicolai Meinshausen.
\newblock Invariant causal prediction for nonlinear models.
\newblock \emph{Journal of Causal Inference}, 6\penalty0 (2), 2018.

\bibitem[Hyttinen et~al.(2012)Hyttinen, Eberhardt, and
  Hoyer]{hyttinen2012learning}
Antti Hyttinen, Frederick Eberhardt, and Patrik~O. Hoyer.
\newblock Learning linear cyclic causal models with latent variables.
\newblock \emph{The Journal of Machine Learning Research}, 13\penalty0
  (109):\penalty0 3387--3439, 2012.

\bibitem[Imbens and Rubin(2015)]{imbens2015causal}
Guido~W Imbens and Donald~B Rubin.
\newblock \emph{Causal inference in statistics, social, and biomedical
  sciences}.
\newblock Cambridge University Press, 2015.

\bibitem[Johnson et~al.(2007)Johnson, Li, and Rabinovic]{johnson2007adjusting}
W~Evan Johnson, Cheng Li, and Ariel Rabinovic.
\newblock Adjusting batch effects in microarray expression data using empirical
  bayes methods.
\newblock \emph{Biostatistics}, 8\penalty0 (1):\penalty0 118--127, 2007.

\bibitem[Judea(2000)]{judea2000causality}
Pearl Judea.
\newblock Causality: models, reasoning, and inference.
\newblock \emph{Cambridge University Press. ISBN 0}, 521\penalty0
  (77362):\penalty0 8, 2000.

\bibitem[Kaplan(2008)]{kaplan2008structural}
David Kaplan.
\newblock \emph{Structural equation modeling: Foundations and extensions},
  volume~10.
\newblock Sage Publications, 2008.

\bibitem[Korb et~al.(2004)Korb, Hope, Nicholson, and Axnick]{korb2004varieties}
Kevin~B Korb, Lucas~R Hope, Ann~E Nicholson, and Karl Axnick.
\newblock Varieties of causal intervention.
\newblock In \emph{Pacific Rim International Conference on Artificial
  Intelligence}, pages 322--331. Springer, 2004.

\bibitem[Kouw et~al.(2016)Kouw, Van Der~Maaten, Krijthe, and
  Loog]{kouw2016feature}
Wouter~M Kouw, Laurens~JP Van Der~Maaten, Jesse~H Krijthe, and Marco Loog.
\newblock Feature-level domain adaptation.
\newblock \emph{The Journal of Machine Learning Research}, 17\penalty0
  (1):\penalty0 5943--5974, 2016.

\bibitem[Pearl(1995)]{pearl1995causal}
Judea Pearl.
\newblock Causal diagrams for empirical research.
\newblock \emph{Biometrika}, 82\penalty0 (4):\penalty0 669--688, 1995.

\bibitem[Pedregosa et~al.(2011)Pedregosa, Varoquaux, Gramfort, Michel, Thirion,
  Grisel, Blondel, Prettenhofer, Weiss, Dubourg, Vanderplas, Passos,
  Cournapeau, Brucher, Perrot, and {{\'E}}douard Duchesnay]{scikit-learn}
Fabian Pedregosa, Ga{{\"e}}l Varoquaux, Alexandre Gramfort, Vincent Michel,
  Bertrand Thirion, Olivier Grisel, Mathieu Blondel, Peter Prettenhofer, Ron
  Weiss, Vincent Dubourg, Jake Vanderplas, Alexandre Passos, David Cournapeau,
  Matthieu Brucher, Matthieu Perrot, and {{\'E}}douard Duchesnay.
\newblock Scikit-learn: Machine learning in python.
\newblock \emph{Journal of Machine Learning Research}, 12\penalty0
  (85):\penalty0 2825--2830, 2011.

\bibitem[Peters et~al.(2016)Peters, B{\"u}hlmann, and
  Meinshausen]{peters2016causal}
Jonas Peters, Peter B{\"u}hlmann, and Nicolai Meinshausen.
\newblock Causal inference by using invariant prediction: identification and
  confidence intervals.
\newblock \emph{Journal of the Royal Statistical Society: Series B (Statistical
  Methodology)}, 78\penalty0 (5):\penalty0 947--1012, 2016.

\bibitem[Pfister et~al.(2019)Pfister, B{\"u}hlmann, and
  Peters]{pfister2019invariant}
Niklas Pfister, Peter B{\"u}hlmann, and Jonas Peters.
\newblock Invariant causal prediction for sequential data.
\newblock \emph{Journal of the American Statistical Association}, 114\penalty0
  (527):\penalty0 1264--1276, 2019.

\bibitem[Pinheiro and Bates(2000)]{pinheiro2000linear}
Jos{\'e}~C Pinheiro and Douglas~M Bates.
\newblock Linear mixed-effects models: basic concepts and examples.
\newblock \emph{Mixed-effects models in S and S-Plus}, pages 3--56, 2000.

\bibitem[{R Core Team}(2017)]{R}
{R Core Team}.
\newblock \emph{R: A Language and Environment for Statistical Computing}.
\newblock R Foundation for Statistical Computing, Vienna, Austria, 2017.
\newblock URL \url{https://www.R-project.org/}.

\bibitem[Rojas-Carulla et~al.(2018)Rojas-Carulla, Sch{\"o}lkopf, Turner, and
  Peters]{rojas2018invariant}
Mateo Rojas-Carulla, Bernhard Sch{\"o}lkopf, Richard Turner, and Jonas Peters.
\newblock Invariant models for causal transfer learning.
\newblock \emph{The Journal of Machine Learning Research}, 19\penalty0
  (1):\penalty0 1309--1342, 2018.

\bibitem[Rothenh{\"a}usler et~al.(2015)Rothenh{\"a}usler, Heinze, Peters, and
  Meinshausen]{rothenhausler2015backshift}
Dominik Rothenh{\"a}usler, Christina Heinze, Jonas Peters, and Nicolai
  Meinshausen.
\newblock Backshift: Learning causal cyclic graphs from unknown shift
  interventions.
\newblock In \emph{Advances in Neural Information Processing Systems}, pages
  1513--1521, 2015.

\bibitem[Rothenh{\"a}usler et~al.(2018)Rothenh{\"a}usler, Meinshausen,
  B{\"u}hlmann, and Peters]{rothenhausler2018anchor}
Dominik Rothenh{\"a}usler, Nicolai Meinshausen, Peter B{\"u}hlmann, and Jonas
  Peters.
\newblock Anchor regression: heterogeneous data meets causality.
\newblock \emph{arXiv preprint arXiv:1801.06229}, 2018.

\bibitem[Rothenh{\"a}usler et~al.(2019)Rothenh{\"a}usler, B{\"u}hlmann,
  Meinshausen, et~al.]{rothenhausler2019causal}
Dominik Rothenh{\"a}usler, Peter B{\"u}hlmann, Nicolai Meinshausen, et~al.
\newblock Causal dantzig: fast inference in linear structural equation models
  with hidden variables under additive interventions.
\newblock \emph{The Annals of Statistics}, 47\penalty0 (3):\penalty0
  1688--1722, 2019.

\bibitem[Schaffter et~al.(2011)Schaffter, Marbach, and
  Floreano]{schaffter2011genenetweaver}
Thomas Schaffter, Daniel Marbach, and Dario Floreano.
\newblock Genenetweaver: in silico benchmark generation and performance
  profiling of network inference methods.
\newblock \emph{Bioinformatics}, 27\penalty0 (16):\penalty0 2263--2270, 2011.

\bibitem[Sinha et~al.(2017)Sinha, Namkoong, and Duchi]{sinha2017certifying}
Aman Sinha, Hongseok Namkoong, and John Duchi.
\newblock Certifying some distributional robustness with principled adversarial
  training.
\newblock \emph{arXiv preprint arXiv:1710.10571}, 2017.

\bibitem[Spirtes et~al.(2000)Spirtes, Glymour, Scheines, and
  Heckerman]{spirtes2000causation}
Peter Spirtes, Clark~N. Glymour, Richard Scheines, and David Heckerman.
\newblock \emph{Causation, prediction, and search}.
\newblock MIT Press, 2000.

\bibitem[Tian and Pearl(2013)]{tian2013causal}
Jin Tian and Judea Pearl.
\newblock Causal discovery from changes.
\newblock \emph{arXiv preprint arXiv:1301.2312}, 2013.

\bibitem[Tibshirani(1996)]{tibshirani1996regression}
Robert Tibshirani.
\newblock Regression shrinkage and selection via the lasso.
\newblock \emph{Journal of the Royal Statistical Society: Series B (Statistical
  Methodology)}, 58\penalty0 (1):\penalty0 267--288, 1996.

\bibitem[Waskom(2021)]{michael_waskom_2017_883859}
Michael~L. Waskom.
\newblock seaborn: statistical data visualization.
\newblock \emph{Journal of Open Source Software}, 6\penalty0 (60):\penalty0
  3021, 2021.

\bibitem[Zou et~al.(2014)Zou, Lippert, Heckerman, Aryee, and
  Listgarten]{zou2014epigenome}
James Zou, Christoph Lippert, David Heckerman, Martin Aryee, and Jennifer
  Listgarten.
\newblock Epigenome-wide association studies without the need for cell-type
  composition.
\newblock \emph{Nature Methods}, 11\penalty0 (3):\penalty0 309--311, 2014.

\end{thebibliography}

\newpage

\onecolumn

\appendix

\section{Supplement on simulations}\label{appendix:simulations}

\subsection{Detailed description of the experimental setting}\label{appendix:simulation-environments}

We compare the following methods: \textbf{1) KL regression} with and without $L^1$-penalty regularization. \textbf{2) Invariant Causal Prediction (ICP)} \citep{peters2016causal} This is a feature selection method, we use the implementation \texttt{ICP} in the R package \texttt{InvariantCausalPrediction}. Given that this method can not handle settings with latent confounders, in our experiments we systematically have a message indicating that \texttt{ICP} does not reject any covariate and redirects to another method in that package under the name \texttt{HiddenICP} \citep{R}. \textbf{3) Anchor Regression} \citep{rothenhausler2018anchor} This method requires fixing a tuning parameter $\gamma$: whenever the estimators obtained using the two extreme choices of $\gamma$ agree (i.e. Anchor stability \citep[Theorem 3, p.15]{rothenhausler2018anchor}), then the choice of $\gamma$ has no impact. This is what we observe in practice in our simulations, hence we report results for the unique anchor regression estimate. \textbf{4) Causal Dantzig} \citep{rothenhausler2019causal} We run experiments with both the regularized and unregularized Causal Dantzig. The regularization parameter is automatically selected by cross-fitting. \textbf{5) ComBat} \citep{johnson2007adjusting}. We use a popular empirical Bayes method for removing batch effects where we consider each environment corresponds to a batch of data. \textbf{6) Averaging the regression coefficients}. For each environment $e \in [E]$, compute the least squares estimator $\hat{\beta}_e$, and then take the average over environments: $\hat{\beta}^{avg} := \frac{1}{E}\sum_{e}\hat{\beta}_e$. This method will serve as a comparison baseline. We also tested Invariant Risk Minimization \citep{arjovsky2019invariant}, but it only performed correctly in low dimension settings ($dim(\Xb) < 10$) and required a time-consuming hyperparameter search.

We generate samples for different environments where the distribution of $\epsilon$ is multivariate Gaussian. The following description corresponds to a default configuration of the parameters, and different simulations analyze the behavior of different estimators by modifying a subset of the parameters along a range of values. Most of the simulations in this section showcase the effect of modifying a specific parameter in the model. We now define a \emph{baseline} configuration for generating data. For each experiment we report, we indicate which of the runs corresponds to this baseline configuration. Our baseline simulation is the following: we generate each environment where the observed covariate dimension is 100, the dimension of the hidden space is 2. We have $\beta^* = (1,1,\dots,1,0,\dots,0)$ where only the first $10$ coefficients are non-zero and $\eta_0 = (0.5,0.5)$. The connectivity matrices $B_{XX}, B_{XH}$ are randomly sampled so that each entry has a binomial distribution: these matrices are sampled once for each simulation, which contains multiple environments. In our baseline simulation, we generate data for $4$ environments, and each environment we generate the same number of samples (set to 5000). Crucially when generating environment data, the only varying parameter across environments is the distribution of $\epsilon_{e,X}, \epsilon_{e,Y}$. Given that these are centered multivariate Gaussians, the varying parameters across environments is the covariance matrix $\Sigma_{e,\epsilon_X}, \Sigma_{e,\epsilon_Y}$ (and we fix $\Sigma_{\epsilon_H} = I_Q$). In the baseline simulation, $\Sigma_{e,\epsilon_X}$ is sampled independently for each environment, and $\Sigma_{e,\epsilon_Y}$ is equal to $1+|Z_e|$ where $Z_e$ are independent standard Gaussian. For any given method, we measure its performance by comparing the mean squared error between the output estimator $\hat{\beta}$ and the invariant $\beta^*$. We report the MSE in a log-scale for better visualization. We repeat each simulation 50 times to provide standard errors, and use enhanced boxplots in \texttt{seaborn} \citep{michael_waskom_2017_883859}. We indicate by a horizontal black line the MSE of the zero vector of coefficients to serve as a benchmark. We ran all our experiments on an Intel Xeon E7 processor (Intel(R) Xeon(R) CPU E5-1650 v4 @ 3.60GHz). A single run of any of these experiments was done in at most 5 minutes.

\subsection{KL regression estimates improve with diverse environments.}\label{appendix:diversity}
As hinted in Proposition~\ref{proposition:invertibility_S}, having diverse covariance matrices $\Sigma_{e,X}$ implies that $S_{\beta}$ is likely to be invertible and well-conditioned, and thus the asymptotic variance of our KL regression estimator will be lower (cf. Proposition~\ref{proposition:distribution-DR}). Following Proposition~\ref{proposition:covariance-parameters}, such environment-wise variability is due to changes in $\Sigma_{e,\epsilon_X}, \Sigma_{e,\epsilon_Y}$. We generate random positive semi-definite matrices $(\Sigma_0, \Sigma_1, \dots, \Sigma_E)$ through the Scikit-Learn \texttt{sklearn.datasets.make\_spd\_matrix} function \citep{scikit-learn}. The simulation analyzes the effects of diversity in $\Sigma_{e,\epsilon_X}$, where we linearly interpolate for $t \in [0,1]$ between the setting where there is only a shared covariance (for $t = 0$), and a setting where each environment has its own independent random sample of $\Sigma_{e,\epsilon_X}$ (for $t=1$, and this correspond to our baseline configuration described in section~\ref{appendix:simulation-environments}) That is, $\Sigma_{e,\epsilon_X} = (1-t)\Sigma_0 + t\Sigma_e$ We report in Figure~\ref{fig:environment-diversity} the results of simulations: all methods improve with the diversity of the environments, but in relative terms within a method's performance KL regression is particularly sensitive to such diversity.

\subsection{Simulations with varying strength of confounding}\label{appendix:varying-confounding}
\begin{figure*}[ht]
\centering
\includegraphics[width=1\linewidth]{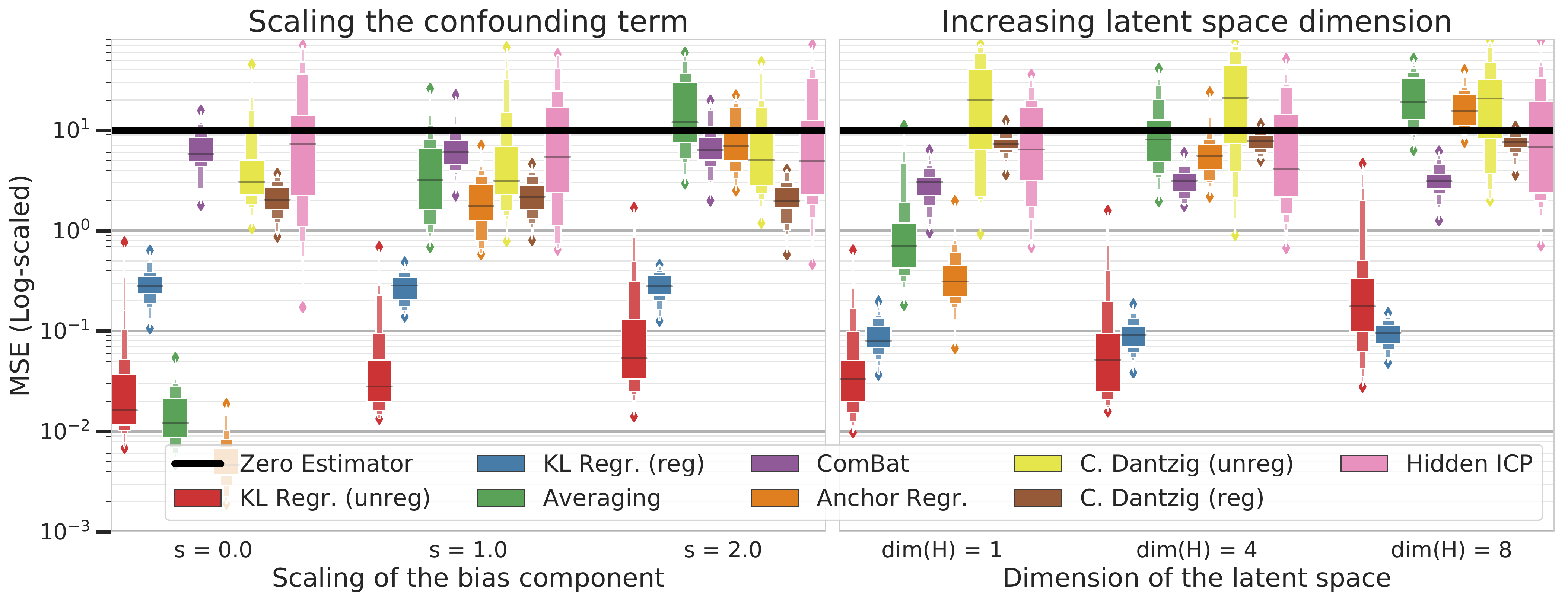}
\caption{\textbf{Varying strength of confounding} Plots of the (log-scaled) MSE of the estimator of the regression parameter (with respect to the true $\beta^*$) as a function of the scaling of $\eta_0$ (left plot), and the dimension of $H$ (right plot). KL regression and regularized Causal Dantzig are the only ones robust to increased confounding, and KL regression systematically recovers $\beta^*$.}
\label{fig:varying-confounding}
\end{figure*}
The latent variables in our model affect both the regression coefficient and the covariance structure of the covariates. We run two experiments to showcase the effect of the confounding on the performance of the estimators. First, we scale the parameter $\eta_0$ by $s$ (which is set to $s=1$ in the baseline simulation) so that environments are generated with $s\eta_0$ instead. We report in Figure~\ref{fig:varying-confounding} (left) the results of this experiment for $s\in \{0,1,2\}$ (where $s=1$ corresponds to our baseline configuration described in section~\ref{appendix:simulation-environments}). Under no confounding, all methods do well (except ComBat, hiddenICP and the unregularized Causal Dantzig which is unstable). With increased confounding, the only methods that do well are regularized Causal Dantzig and ours, which perfectly recovers $\beta^*$ (repeating this experiment we observed that regularized Causal Dantzig is slightly more variable, but usually is the best alternative to KL regression). Under a sparser setting for $\beta^*$, we observed that Causal Dantzig substantially improved and matches KL regression performance, regardless of the confounding strength. In the second experiment, we increase the dimension $Q$ of the latent space $H$ (so that the baseline simulation is $Q = 2$ as described in section~\ref{appendix:simulation-environments}). We report in Figure~\ref{fig:varying-confounding} (right) the results of this experiment for $Q\in \{1,4,8\}$. The results observed match those of the first experiment: KL regression consistently ranks among the best methods.

We also ran experiments comparing KL regression against Invariant Causal Prediction (ICP). The outcome of such simulations was consistently the same: ICP is able to detect the presence of latent confounders in the model and outputs a message indicating that none of the variables had been rejected, and redirects the user to the \texttt{hiddenICP} command in the package, which we ran and reported the corresponding results. 

\subsection{Effect of changes in sparsity of the invariant vector}\label{appendix:sparsity}
\begin{figure*}[ht]
\centering
\includegraphics[width=0.8\linewidth]{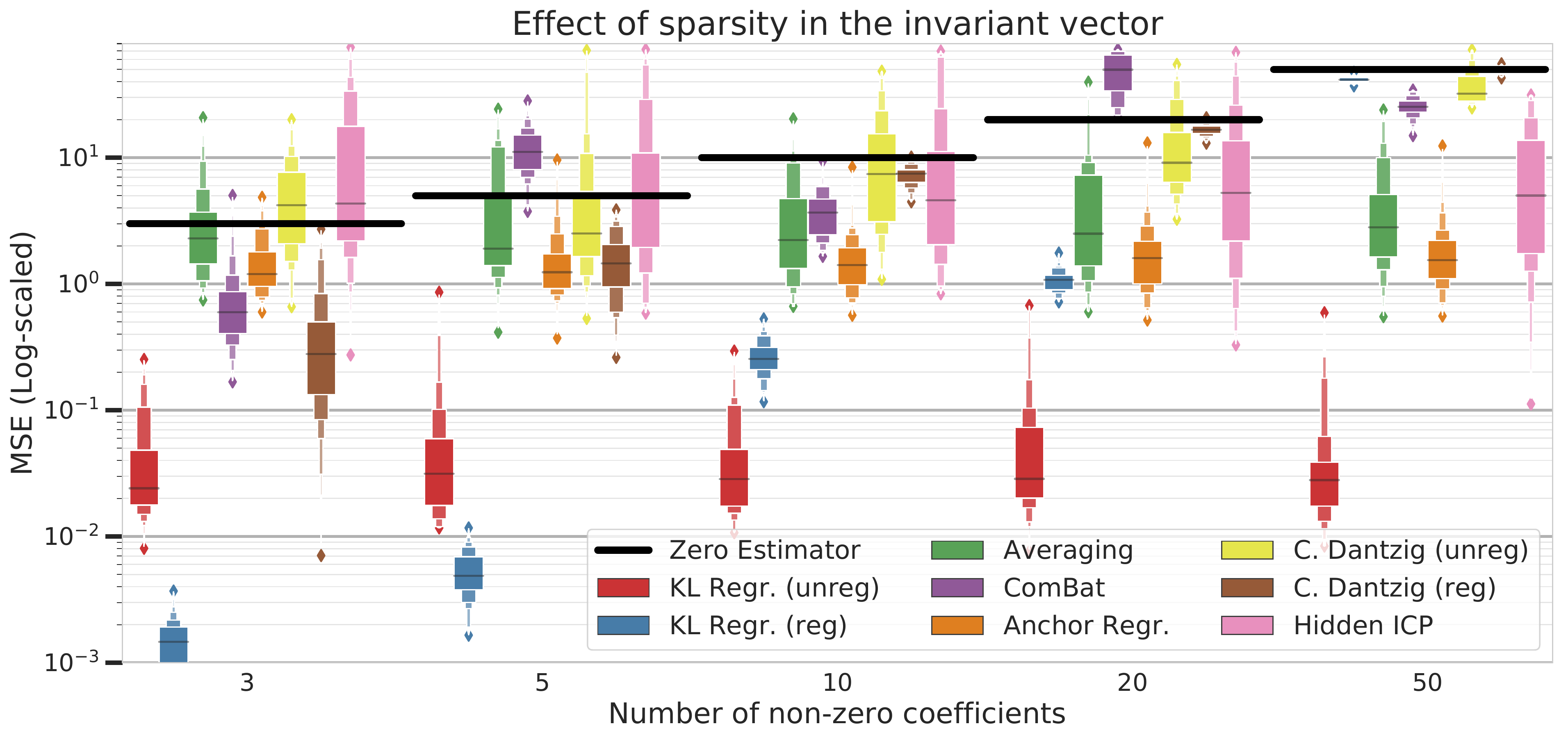}
\caption{\textbf{Changing the sparsity level of $\beta^*$.} Plots of the (log-scaled) MSE of the estimator of the regression parameter (with respect to the true $\beta^*$) as a function of the sparsity level of $\beta^*$. KL regression does not require any sparsity assumption, and outperforms competing methods (which some do require sparsity).}
\label{fig:sparsity}
\end{figure*}
Unregularized KL regression does not require $\beta^*$ to be sparse in order to work. Most of the other methods do not need it either, except regularized Causal Dantzig, which substantially improves over unregularized Causal Dantzig. This sparsity-inducing regularization version of Causal Dantzig and of KL regression perform well in most settings generated based on our baseline configuration described in section~\ref{appendix:simulation-environments} where $\beta^*$ is sparse. We show in this simulation that by increasing the number of non-zero coefficients of $\beta^*$, unregularized KL regression is the only method that is still exactly recovers the invariant vector. We report the results of the simulation in Figure~\ref{fig:sparsity}, where we the number $d_0$ of non-null coefficients in $\beta^*$ is in $\{3,5,10,20,50\}$ (where $d_0 = 10$ would correspond to the baseline).

\subsection{Splitting the data from one environment to increase the number of environments does not help KL regression}\label{appendix:split}
\begin{figure*}[ht]
\centering
\includegraphics[width=0.65\linewidth]{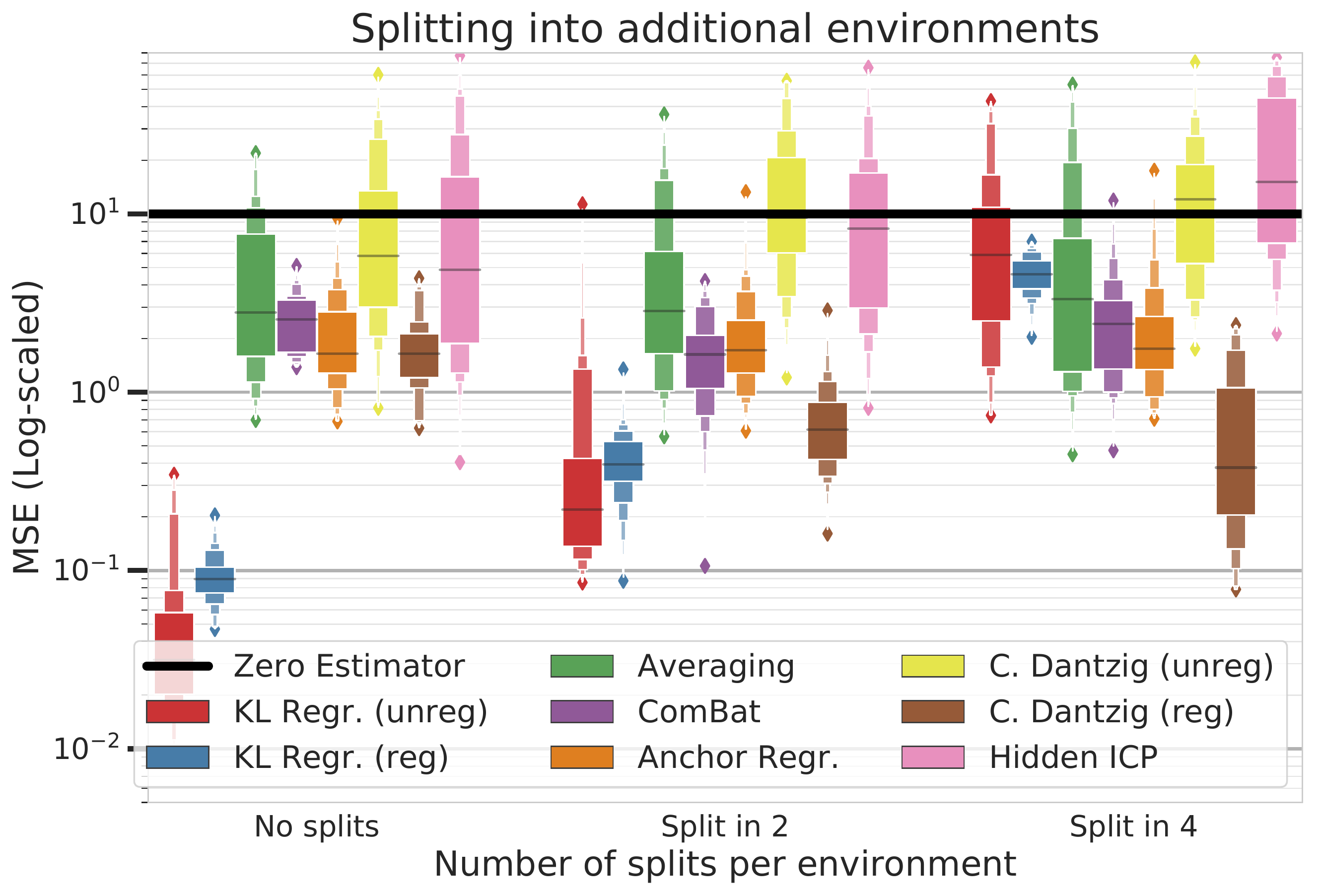}
\caption{\textbf{Creating additional environments by splitting existing ones} Plots of the (log-scaled) MSE of the estimator of the regression parameter (with respect to the true $\beta^*$) by comparing the baseline setting (cf. section~\ref{appendix:simulation-environments}) with a dataset where environments are split in two. Performance of KL regression degrades due to the splits.}
\label{fig:split}
\end{figure*}

We generate different simulations by randomly splitting the data from each environment in order to create additional environments. We compare the baseline configuration in section~\ref{appendix:simulation-environments} to the setting where each environment is split in two different datasets. We report the results in Figure~\ref{fig:split}. The performance of KL regression worsens as a consequence of the split, although the amount of information provided by the samples is the same. The newly generated environments through splits only make the covariance estimation noisier, and do not provide extra information as two random partitions from the same initial dataset share the same covariance structure, hence do not contribute to reducing the variance of the KL regression estimator. Other methods suffer less from splitting.

\subsection{KL regression does not require Gaussianity.}\label{appendix:student}
\begin{figure*}[ht]
\centering
\includegraphics[width=0.8\linewidth]{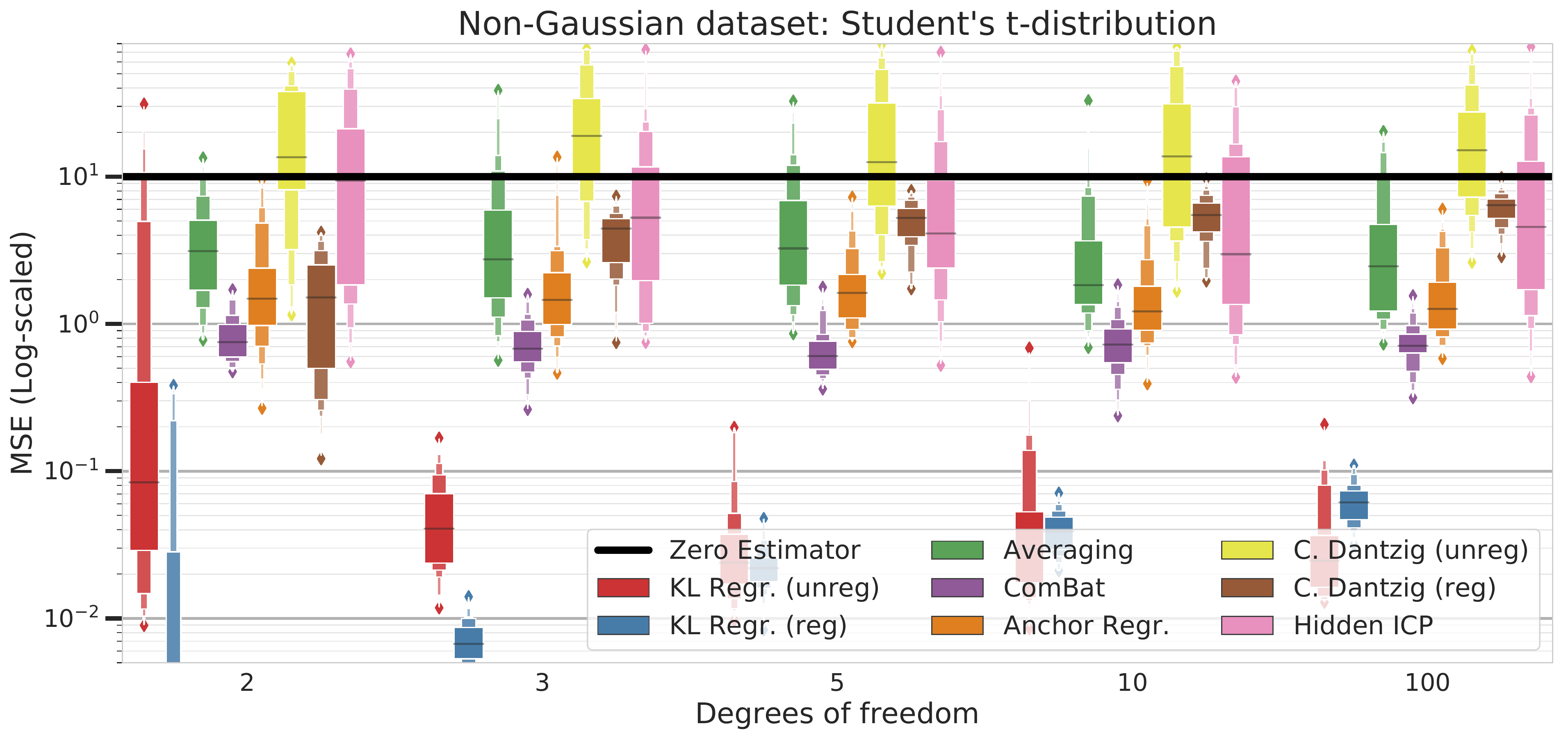}
\caption{\textbf{Generating $\epsilon$ from Student's t-distribution} Plots of the (log-scaled) MSE of the estimator of the regression parameter (with respect to the true $\beta^*$) where the difference with respect to the baseline (cf. section~\ref{appendix:simulation-environments}) is that samples are generated from a t-distribution with different degrees of freedom $\nu$. Smaller values imply thicker tails of the data-generating distribution. Aside from some instability for low values of $\nu$, the outcomes of this simulation is generally consistent with previous ones.}
\label{fig:student}
\end{figure*}

Although the loss used to define KL regression can be interpreted as a KL regression between two multivariate Gaussians, all of our results hold regardless of the actual distribution of the random vector $\epsilon$ defining the SEM. It is only based on the linearity assumption when defining the SEM (cf. equation~\ref{equation-linear-sem}) and our model of environments that we can recover $\beta^*$. We report in Figure~\ref{fig:student} simulations where data is generated from a heavy-tail distribution. Aside from higher variance, we indeed do not observe a decrease in performance of KL regression.

\subsection{DREAM dataset analysis}\label{appendix:DREAM}

We compare KL regression and Causal Dantzig on two DREAM challenge datasets. We compare the regularized methods that allow to select features and not just estimating the vector of coefficients. The task at hand is to recover the gene regulatory network, for which a ground truth is known. The specific coefficient values do not yield any additional information besides being non-zero. We also compare our methods to the leaderboard of the two competitions.

\subsubsection{DREAM4 Dataset}
The DREAM4 10-node in silico dataset simulates gene regulatory network expressions from 10 different yeast genes under different types of perturbations \citep{schaffter2011genenetweaver}. Although synthetic data, it is a widely-used and challenging testing ground where the causal mechanisms are known. The DREAM4 dataset \citep{schaffter2011genenetweaver} contains gene expression simulated data based on curated and expert knowledge of the gene regulatory networks in yeast. The dataset consists on time series corresponding to the dynamic behavior of the gene expression under several types of perturbation in the genes. We use the 10-node in silico dataset. This dataset contains 5 different sets of simulations, corresponding to 5 distinct independent regulatory networks. Each of this distinct datasets contains a matrix of gene expression, where samples correspond to different experiments/time points of an experiment. We generate 6 environments, consisting on 5 separate sets, each of them based on a time series corresponding to a dynamic on/off perturbation on different parts of the network. The 6th set that corresponds to the knock-down of genes one at a time. We have access to a ground truth connectivity matrix reflecting the underlying ground truth of the regulatory network. Because of the presence of loops in the network, several of the assumptions that ICP, Causal Dantzig and KL regression are based upon are not satisfied. In particular, using a Bonferroni correction to control the number of false discoveries is not necessarily successful, as the p-values may not be uniformly distributed. Also, for each perturbation there is only one sample, which forces us to combine several perturbation within the same environment. 

We consider the ground truth to be the undirected graph indicating a connection between the expression of two genes, and we loop over the different genes taking the gene expression of each of them as our response variable. We compare regularized Causal Dantzig and Lasso KL regression for the task of feature selection. We report in Table~\ref{table:DREAM} the metric area under the precision-recall curve, which was used in the DREAM competition and allows us to benchmark our model against competing methods. The leaderboard for this competition can be found at https://www.synapse.org/\#!Synapse:syn3049712/wiki/74631. Lasso KL regression performs decently, ranking in the top 40\% of the submissions. 

\subsubsection{DREAM5 Dataset}

The DREAM5 dataset contains 4 different datasets corresponding to gene expression measurements. Dataset 1 is generated similarly to that of DREAM4, except that some perturbations have several experimental replicates. Also, instead of focusing on the regulatory interaction between 10 genes, this dataset has a more realistic configuration where more than 800 genes are studied: among those, 185 genes are potential candidates for regulating the other genes. That is, 185 transcription factors can potentially regulate any of the 800 potential target genes. This problem has a higher dimension than DREAM4, but we still expect a few number of transcription factors to regulate any given gene. Therefore the regularized Causal Dantzig and the Lasso KL regression are better suited for discoveries in this dataset. We run simulations by looping over the potential target genes (and actually focus only on the non-transcription factor genes) and obtain a regularization path by varying in both techniques the regularization strength. We compute the precision/recall curve, and Lasso KL Regression \emph{outperforms every other competing method in the DREAM challenge}. We report in Table~\ref{table:DREAM} the metric area under the precision-recall curve, which was used in the DREAM competition and allows us to benchmark our model against competing methods. The leaderboard for this challenge can be found at https://www.synapse.org/\#!Synapse:syn2787209/wiki/70352. All these simulations can be reproduced in with the code we provide.

Unfortunately, this simulation revealed a weakness of KL regression. Although we regularize the loss in a similar way to Lasso regression, we still require the environment-wise covariance matrices to be positive definite. This is not the case if the number of samples is smaller than the dimension of the covariate space, and in the remaining datasets of the DREAM5 challenge there were not enough samples to generate several environments. However, the emerging field of genome editing CRISPR-Cas9 genome-wide screens \citep{gasperini2019genome}, where single cell sequencing is done at a very large scale may benefit from KL regression as the perturbations induced by the genome editing knock-down interventions are a good framework for a multiple environment causal inference analysis.

\section{Analogy with linear regression}\label{appendix:regression-analogy}

In linear regression, one individual sample consists of covariates $\Xb = (X_d)_{1\leq d \leq D}\in \mathbb{R}^D$ and a response $Y$. The internal relationship between those is modeled through a regression function which is a linear combination of the covariates for some estimated regression coefficient $\hat{\beta}\in \mathbb{R}^D$:
\begin{align}\label{equation:linear}
\hat{Y} = \hat{Y}(\hat{\beta}) = \hat{\beta}^T\Xb = \sum_{1\leq d \leq D} \hat{\beta}_d X_d
\end{align} 
We choose $\hat{\beta}$ so that we minimize the discrepancy between $\hat{Y}$ and the response $Y$ in the sample (usually by minimizing sum of squares). Linear regression tries to capture an underlying structure of each sample via the estimated coefficient $\hat{\beta}$ (that is, the linear relationship between the covariates and the response), and requires multiple observations sharing this common structure to output a good estimate. Our intuition for defining KL regression follows those lines. We take a step back and consider each of the environments in the set $\mathcal{P}_E := \{\pi_e\}_{1\leq e \leq E}$ as a ``meta-sample''. Given the linear SEM structure, we focus on second-order moments of the distributions we consider. Therefore for each $\pi_e$ we focus on its covariance structure, and in particular we decompose it so that the \emph{regression function} mapping $\Pi$ separates the sample-wise component of the covariance of $\pi_e$, from the shared invariant $\theta$. This parameter $\theta$ will reflect a shared intrinsic quantity of the distributions in $\mathcal{P}_E$, and will be the target object to estimate (the equivalent of the regression coefficient $\beta$). In particular, for a given estimate $\hat{\theta}$ that we compute based on data, for $\pi_e \in \mathcal{P}_E$, we denote by $\hat{\pi}_e = \hat{\pi}_e(\hat{\theta}) := \Pi(\Sigma_e, \sigma_e^2, \hat{\theta})$ (equivalent of $\hat{Y}$) where $\Sigma_e, \sigma_e^2$ are the individual sample regressors. Analogously if we were to write the regression function in the usual sense, for a sample $(\Xb, Y)$ and a parameter $\beta$, we would get $\Pi\big((\Xb, Y), \beta \big) = \big(\Xb, \beta^T\Xb\big)$.

Finally, we use some measure of dissimilarity between probability distributions $\rho: \mathcal{P}^{all}\times\mathcal{P}^{all}\rightarrow \mathbb{R}_+$ such that $\rho(\pi, \pi') = 0$ if and only if $\text{Cov}(\pi) =  \text{Cov}(\pi')$ (the equivalent of the squared difference in linear regression). As second-order statistics characterize multivariate Gaussian distributions, we use the Kullback-Leibler divergence as a proxy for $\rho$. then solves the following minimization problem:
\begin{align}\label{loss-generic}
    & \hat{\theta} := \argmin_{\theta \in \Theta} \mathcal{L}(\theta)
    \\ & \mathcal{L}(\theta) := \sum_{e\in[E]} \rho(\pi_e, \hat{\pi}_e(\theta))\nonumber
\end{align}{}
Linear regression in turn is characterized by $\rho\big((\Xb, Y),(\Xb, \beta^T\Xb)\big) = (Y-\beta^T\Xb)^2$. Now the question is to analyze what is actually estimated through the \eqref{loss-generic} objective. This is where we need to take into account the data-generating model. In particular, we expect that $\hat{\theta}$ is estimating some value $\theta^*$ that is constant across the environments, which is the invariant object of interest. We will therefore need to adapt the parametrization in $\theta$ of the distribution regression function to the model we consider for how $(\pi_e)_e$ are generated. We now provide a closed-form solution for problem~\ref{loss-eq-simple} in the particular context of our linear SEM (cf. equation~\ref{equation-linear-sem}).

\begin{proposition}
The solution to the minimization problem~\eqref{loss-eq-simple} has a closed-form solution given by:
\begin{align*}
    \tilde{\theta} = \Big( \sum_{e \in [E]}\frac{\Sigma_{e,X}}{\sigma_e^2}\Big)^{-1}\Big(\sum_{e \in [E]} \frac{\Sigma_{e,X}\beta_e}{\sigma_e^2}\Big)
\end{align*}{}
\end{proposition}{}

In practice, the elements of the triplet $(\beta_e, \Sigma_{e,X}, \sigma_e^2)$ are estimated based on i.i.d. samples from $\pi_e$: i.e. our samples for KL regression are given by the estimates $\hat{\Sigma}_e^{joint}$ of $\Sigma_e^{joint}$ and the subsequent transformations into $(\hat{\beta}_e, \hat{\Sigma}_{e,X}, \hat{\sigma}_e^2)$. That is, the actual estimator $\hat{\theta}$ is given by:
\begin{align*}
    \hat{\theta} = \Big( \sum_{e \in [E]}\frac{\hat{\Sigma}_{e,X}}{\hat{\sigma}_e^2}\Big)^{-1}\Big(\sum_{e \in [E]} \frac{\hat{\Sigma}_{e,X} \hat{\beta}_e}{\hat{\sigma}_e^2}\Big)
\end{align*}{}

Finally, the analogy between Lasso KL regression and usual Lasso regression is immediate, by considering an $L^1$-norm penalty on the regression coefficient in the minimization problem.

\section{Proofs}

\subsection{Proof of Proposition~\ref{proposition:covariance-parameters}}

\begin{proof}\label{proof:proposition:covariance-parameters}
We prove here Proposition~\ref{proposition:covariance-parameters}. We want to compute
\begin{align*}
    & \text{Cov} \begin{pmatrix}{}
    \Xb_e \\ Y_e \\ \Hb_e
    \end{pmatrix} = (I-\mathbf{B})^{-1}\text{Cov}(\epsilon_e)\big((I - \mathbf{B})^{-1}\big)^T
\end{align*}{}
where $\epsilon_e := (\epsilon_{e,X}, \epsilon_{e,Y}, \epsilon_H)$. We have that, with $C := (I -B_{XX})^{-1}$:
\begin{align*}
    & (I - \mathbf{B}) = \begin{pmatrix}
    I_D - B_{XX} & 0 & -B_{XH}
    \\ -\beta^{*T} & 1 & -\eta_0^T
    \\ 0 & 0 & I_Q
    \end{pmatrix}
    \\ & (I - \mathbf{B})^{-1} = \begin{pmatrix}
    (I_D-B_{XX})^{-1} & 0 & (I_D-B_{XX})^{-1}B_{XH}
    \\ \beta^{*T}(I_D-B_{XX})^{-1} & 1 & \eta_0^T + \beta^{*T}(I_D-B_{XX})^{-1}B_{XH}
    \\ 0 & 0 & I_Q
    \end{pmatrix} = \begin{pmatrix}
    C & 0 & CB_{XH}
    \\ \beta^{*T}C & 1 & \eta_0^T + \beta^{*T}CB_{XH}
    \\ 0 & 0 & I_Q
    \end{pmatrix}
    \\& (I - \mathbf{B})^{-1}\text{Cov}(\epsilon_e) = \begin{pmatrix}
    C\Sigma_{e, \epsilon_X} & 0 & CB_{XH}\Sigma_{\epsilon_H}
    \\ \beta^{*T}C\Sigma_{e, \epsilon_X} & \Sigma_{e,\epsilon_Y} & \eta_0^T\Sigma_{\epsilon_H} + \beta^{*T}CB_{XH}\Sigma_{\epsilon_H}
    \\ 0 & 0 & \Sigma_{\epsilon_H}
    \end{pmatrix}
\end{align*}
We then have the following expressions for the covariance of $\Xb_e$ and $(\Xb_e,Y_e)$ by taking the $D+1$ principal sub-matrix of $(I-\mathbf{B})^{-1}\text{Cov}(\epsilon_e)\big((I - \mathbf{B})^{-1}\big)^T$.
\begin{align*}
    \begin{cases}
        \text{Cov}(\Xb_e) = C\Sigma_{e, \epsilon_X}C^T + CB_{XH}\Sigma_{\epsilon_H}B_{XH}^TC^T
        \\ \text{Cov}(\Xb_e, Y_e) = C\Sigma_{e, \epsilon_X}C^T\beta^* + CB_{XH}\Sigma_{\epsilon_H}B_{XH}^TC^T\beta^* + CB_{XH}\Sigma_{\epsilon_H}\eta_0
    \end{cases}{}
\end{align*}{}
Therefore we get that 
\begin{align*}
    \text{Cov}(\Xb_e) & = C\Big(\Sigma_{e,\epsilon_{X}} + B_{XH}\Sigma_{\epsilon_H}B_{XH}^T\Big)C^T
    \\ \text{Cov}(\Xb_e)^{-1}\text{Cov}(\Xb_e, Y_e) & = \beta^* + \text{Cov}(\Xb_e)^{-1}CB_{XH}\Sigma_{\epsilon_H}\eta_0 = \beta^* + \text{Cov}(\Xb_e)^{-1}\eta^*
\end{align*}{}
where $\eta^*= CB_{XH}\Sigma_{\epsilon_H}\eta_0$ does not depend on the environment.
\end{proof}{}

\subsection{Proof of Proposition~\ref{proposition:solution-DR}}

\begin{proof}\label{proof:proposition:solution-DR}
We prove here Proposition~\ref{proposition:solution-DR}. We drop the subscript $X$ from the covariance matrix $\Sigma_{e,X}$ for notation simplicity. Consider the KL regression loss, and our choice of parametrizing the covariance matrix $\Pi\big((\Sigma, \sigma^2),\theta)\big)$ with $\theta = \beta + \Sigma_e^{-1}\eta$. Given $E$ different environments, we get:
\begin{align*}
    \mathcal{L}(\beta, \eta) = & \sum_{e\in [E]} \frac{\big(\beta_e-(\beta + \Sigma_e^{-1}\eta)\big)^T \Sigma_e \big(\beta_e - (\beta + \Sigma_e^{-1}\eta)\big)}{2\sigma_e^2}
\end{align*}
Our estimates $(\tilde{\beta}^{KL}, \tilde{\eta})$ are the solutions to the minimization problem
\begin{equation*}
    \tilde{\beta}^{KL}, \tilde{\eta} = \argmin_{\beta, \eta} \mathcal{L}(\beta,\eta) 
\end{equation*}{}
Given that the problem is convex, continuously differentiable in both arguments, we compute the gradients of the loss in order to find a minimizer.
\begin{align*}
    \mathcal{L}(\beta, \eta) & = \sum_{e\in [E]} \frac{\big(\beta-(\beta_e- \Sigma_e^{-1}\eta)\big)^T \Sigma_e \big(\beta-(\beta_e- \Sigma_e^{-1}\eta)\big)}{2\sigma_e^2}
    \\ \nabla_{\beta}\mathcal{L}(\beta, \eta) & = \sum_{e\in [E]}\frac{1}{\sigma_e^2}\big(\Sigma_e \beta - \Sigma_e(\beta_e - \Sigma_e^{-1}\eta)\big)
    \\ & = \Big(\sum_{e\in [E]}\frac{\Sigma_e}{\sigma_e^2}\Big)\beta - \Big(\sum_{e\in [E]}\frac{\Sigma_e\beta_e}{\sigma_e^2}\Big) + \Big(\sum_{e\in [E]}\frac{1}{\sigma_e^2}\Big)\eta
\end{align*}
and we have
\begin{align*}
   \mathcal{L}(\beta, \eta) & = \sum_{e\in [E]}
   \frac{\big(\Sigma_e^{-1}\eta- (\beta_e - \beta)\big)^T \Sigma_e \big(\Sigma_e^{-1}\eta- (\beta_e - \beta)\big)}{2\sigma_e^2}
   \\ \nabla_{\eta}\mathcal{L}(\beta, \eta) & =  \sum_{e\in [E]}\frac{1}{\sigma_e^2}\big(\Sigma_e^{-1}\eta- (\beta_e - \beta)\big)
   \\ & =  \Big(\sum_{e\in [E]}\frac{\Sigma_e^{-1}}{\sigma_e^2}\Big)\eta- \Big(\sum_{e\in [E]}\frac{\beta_e}{\sigma_e^2}\Big) + \Big(\sum_{e\in [E]}\frac{1}{\sigma_e^2}\Big)\beta
\end{align*}
Therefore setting these derivatives to $0$ we get that:
\begin{align*}
\begin{cases}
    & \nabla_{\beta}\mathcal{L}(\tilde{\beta}^{KL}, \tilde{\eta}) = 0 
    \\ & \nabla_{\eta}\mathcal{L}(\tilde{\beta}^{KL}, \tilde{\eta}) = 0
\end{cases}{} \qquad \Longleftrightarrow
\begin{cases}
    \Big(\sum_{e\in [E]}\frac{\Sigma_e}{\sigma_e^2}\Big)\tilde{\beta}^{KL}  =  \Big(\sum_{e\in [E]}\frac{\Sigma_e\beta_e}{\sigma_e^2}\Big) - \Big(\sum_{e\in [E]}\frac{1}{\sigma_e^2}\Big)\tilde{\eta}
    \\  \Big(\sum_{e\in [E]}\frac{\Sigma_e^{-1}}{\sigma_e^2}\Big)\tilde{\eta}= \Big(\sum_{e\in [E]}\frac{\beta_e}{\sigma_e^2}\Big) - \Big(\sum_{e\in [E]}\frac{1}{\sigma_e^2}\Big)\tilde{\beta}^{KL}
\end{cases}{}
\end{align*}{}
After rearranging the expressions we get that the solution $\tilde{\beta}^{KL}$ satisfies the equation:
\begin{align*}
    & \Bigg(\Big(\sum_{e\in [E]}\frac{\Sigma_e^{-1}}{\sigma_e^2}\Big)\Big(\sum_{e\in [E]}\frac{\Sigma_e}{\sigma_e^2}\Big) - \Big(\sum_{e\in [E]}\frac{1}{\sigma_e^2}\Big)^2I_D\Bigg)\tilde{\beta}^{KL} =  \Big(\sum_{e\in [E]}\frac{\Sigma_e^{-1}}{\sigma_e^2}\Big)\Big(\sum_{e\in [E]}\frac{\Sigma_e\beta_e}{\sigma_e^2}\Big) - \Big(\sum_{e\in [E]}\frac{1}{\sigma_e^2}\Big)\Big(\sum_{e\in [E]}\frac{\beta_e}{\sigma_e^2}\Big) 
\end{align*}{}
Define the scaling matrix $S_{\beta}$ by
\begin{align}
    S_{\beta}:= \Big(\sum_{e\in [E]}\frac{\Sigma_e^{-1}}{\sigma_e^2}\Big)\Big(\sum_{e\in [E]}\frac{\Sigma_e}{\sigma_e^2}\Big) - \Big(\sum_{e\in [E]}\frac{1}{\sigma_e^2}\Big)^2I_D
\end{align}
So that, assuming it is invertible, then we get
\begin{align*}
    \tilde{\beta}^{KL} =& S_{\beta}^{-1}\Bigg(\Big(\sum_{e\in [E]}\frac{\Sigma_e^{-1}}{\sigma_e^2}\Big)\Big(\sum_{e\in [E]}\frac{\Sigma_e\beta_e}{\sigma_e^2}\Big)- \Big(\sum_{e\in [E]}\frac{1}{\sigma_e^2}\Big)\Big(\sum_{e\in [E]}\frac{\beta_e}{\sigma_e^2}\Big)\Bigg)
\end{align*}{}

The solution $\tilde{\beta}^{KL}$ to the optimization problem is designed to estimate $\beta^*$. We now verify that if our model of environments based on shift interventions is correct, then we recover the invariant $\beta^*$. Assume that our distributional samples follow the model, so that $\beta_e = \beta^* + \Sigma_e^{-1}\eta^*$. By plugging this into the formula of the estimator $\tilde{\beta}^{KL}$ we get:
\begin{align*}
    \tilde{\beta}^{KL} =& S_{\beta}^{-1}\Bigg(\Big(\sum_{e\in [E]}\frac{\Sigma_e^{-1}}{\sigma_e^2}\Big)\Big(\sum_{e\in [E]}\frac{\Sigma_e\beta_e}{\sigma_e^2}\Big)- \Big(\sum_{e\in [E]}\frac{1}{\sigma_e^2}\Big)\Big(\sum_{e\in [E]}\frac{\beta_e}{\sigma_e^2}\Big)\Bigg)
    \\ = & S_{\beta}^{-1}\Bigg(\Big(\sum_{e\in [E]}\frac{\Sigma_e^{-1}}{\sigma_e^2}\Big)\Big(\sum_{e\in [E]}\frac{\Sigma_e(\beta^* + \Sigma_e^{-1}\eta^*)}{\sigma_e^2}\Big) - \Big(\sum_{e\in [E]}\frac{1}{\sigma_e^2}\Big)\Big(\sum_{e\in [E]}\frac{(\beta^* + \Sigma_e^{-1}\eta^*)}{\sigma_e^2}\Big)\Bigg)
    \\ = & S_{\beta}^{-1}\Bigg(S_{\beta}\beta^* + \Big(\sum_{e\in [E]}\frac{\Sigma_e^{-1}}{\sigma_e^2}\Big)\Big(\sum_{e\in [E]}\frac{1}{\sigma_e^2}\Big)\eta^*- \Big(\sum_{e\in [E]}\frac{1}{\sigma_e^2}\Big)\Big(\sum_{e\in [E]}\frac{\Sigma_e^{-1}}{\sigma_e^2}\Big)\eta^*\Bigg)
    \\ = & \beta^*
\end{align*}{}
\end{proof}

Hence the result.

\subsection{Proof of Proposition~\ref{proposition:invertibility_S}}

\begin{proof}\label{proof:proposition:invertibility_S}
For completeness, we transcribe the proof from \cite{bhagwat_subramanian_1978} using our notation to provide a proof for Proposition~\ref{proposition:invertibility_S}. Again, we drop the subscript $X$ from the covariance matrix $\Sigma_{e,X}$ for notation simplicity. We assume that $\Sigma_e$ is positive definite for every environment $e\in [E]$. Define
\begin{align*}
    A := & \Big(\sum_{e\in [E]}\frac{\Sigma_e}{\sigma_e^2}\Big) - \Big(\sum_{e\in [E]}\frac{1}{\sigma_e^2}\Big)^2\Big(\sum_{e\in [E]}\frac{\Sigma_e^{-1}}{\sigma_e^2}\Big)^{-1}
    \\ B:= & \Big(\sum_{e\in [E]}\frac{\Sigma_e^{-1}}{\sigma_e^2}\Big)A\Big(\sum_{e\in [E]}\frac{\Sigma_e^{-1}}{\sigma_e^2}\Big)
    \\ = & \Big(\sum_{e\in [E]}\frac{\Sigma_e^{-1}}{\sigma_e^2}\Big)\Big(\sum_{e\in [E]}\frac{\Sigma_e}{\sigma_e^2}\Big)\Big(\sum_{e\in [E]}\frac{\Sigma_e^{-1}}{\sigma_e^2}\Big)  -  \Big(\sum_{e\in [E]}\frac{1}{\sigma_e^2}\Big)^2\Big(\sum_{e\in [E]}\frac{\Sigma_e^{-1}}{\sigma_e^2}\Big) 
    \\ D_f := & P_f\frac{\Sigma_f}{\sigma_f^2} P_f
\end{align*}{}
where 
\[
P_f := \sum_{e\in [E]}\frac{\Sigma_e^{-1}}{\sigma_e^2} - (\sum_{e\in [E]}\frac{1}{\sigma_e^2})\Sigma_f^{-1}
\]
$P_f$ is symmetric, and given that $\Sigma_f$ is positive definite we have that $D_f$ is positive semi-definite (indeed matrix congruence preserves the positive definite character of the matrix when all matrices are invertible, but here $P_f$ is not necessarily invertible). Now, because positive semi-definite symmetric matrices form a convex cone, we get that the sum of the $D_f$ is also positive semi-definite:
\begin{align*}
    \sum_{f \in [E]} D_f = & \sum_{f \in [E]} P_f\frac{\Sigma_f}{\sigma_f^2} P_f
    \\ = & \Big(\sum_{e\in [E]}\frac{\Sigma_e^{-1}}{\sigma_e^2}\Big)\Big(\sum_{f\in [E]}\frac{\Sigma_f}{\sigma_f^2}\Big)\Big(\sum_{e\in [E]}\frac{\Sigma_e^{-1}}{\sigma_e^2}\Big)  - \Big(\sum_{e\in [E]}\frac{1}{\sigma_e^2}\Big)\Big(\sum_{f\in [E]}\Sigma_f^{-1}\frac{\Sigma_f}{\sigma_f^2}\Big)\Big(\sum_{e\in [E]}\frac{\Sigma_e^{-1}}{\sigma_e^2}\Big)
    \\ &\qquad - \Big(\sum_{e\in [E]}\frac{\Sigma_e^{-1}}{\sigma_e^2}\Big)\Big(\sum_{f\in [E]}\frac{\Sigma_f}{\sigma_f^2}(\sum_{e\in [E]}\frac{1}{\sigma_e^2})\Sigma_f^{-1}\Big) + \Big(\sum_{e\in [E]}\frac{1}{\sigma_e^2}\Big)^2\Big(\sum_{f\in [E]}\Sigma_f^{-1}\frac{\Sigma_f}{\sigma_f^2}\Sigma_f^{-1}\Big)
    \\ = & \Big(\sum_{e\in [E]}\frac{\Sigma_e^{-1}}{\sigma_e^2}\Big)\Big(\sum_{f\in [E]}\frac{\Sigma_f}{\sigma_f^2}\Big)\Big(\sum_{e\in [E]}\frac{\Sigma_e^{-1}}{\sigma_e^2}\Big) - \Big(\sum_{e\in [E]}\frac{1}{\sigma_e^2}\Big)^2\Big(\sum_{e\in [E]}\frac{\Sigma_e^{-1}}{\sigma_e^2}\Big) 
    \\ = & B
\end{align*}{}
Hence $B$ is positive semi-definite, and thus $A$ is positive semi-definite too. As $\Big(\sum_{e\in [E]}\frac{\Sigma_e^{-1}}{\sigma_e^2}\Big)$ is positive definite, we get that $S_{\beta}$ is invertible as long as $B$ is invertible. But, for $B$ not to be positive definite, we need all the $P_f$ matrices not to be invertible either, hence the result.
\end{proof}{}

\subsection{Proof of Proposition~\ref{proposition:robustness}}

\begin{proof}\label{proof:proposition:robustness}
We have that
\begin{align*}
    S_{\beta}\tilde{\beta}^{KL} =& \Big(\sum_{e\in [E]}\frac{\Sigma_e^{-1}}{\sigma_e^2}\Big)\Big(\sum_{e\in [E]}\frac{\Sigma_e\beta_e}{\sigma_e^2}\Big)- \Big(\sum_{e\in [E]}\frac{1}{\sigma_e^2}\Big)\Big(\sum_{e\in [E]}\frac{\beta_e}{\sigma_e^2}\Big)
    \\ = & \Big(\sum_{e\in [E]}\frac{\Sigma_e^{-1}}{\sigma_e^2}\Big)\Big(\sum_{e\in [E]}\frac{\Sigma_e(\beta^* + \Sigma_e^{-1}\eta^* + \Sigma_e^{-1}\delta_e)}{\sigma_e^2}\Big) - \Big(\sum_{e\in [E]}\frac{1}{\sigma_e^2}\Big)\Big(\sum_{e\in [E]}\frac{(\beta^* + \Sigma_e^{-1}\eta^*+ \Sigma_e^{-1}\delta_e)}{\sigma_e^2}\Big)
    \\ = & S_{\beta}\beta^* + \Big(\sum_{e\in [E]}\frac{\Sigma_e^{-1}}{\sigma_e^2}\Big)\Big(\sum_{e\in [E]}\frac{\delta_e}{\sigma_e^2}\Big)- \Big(\sum_{e\in [E]}\frac{1}{\sigma_e^2}\Big)\Big(\sum_{e\in [E]}\frac{\Sigma_e^{-1}\delta_e}{\sigma_e^2}\Big)
\end{align*}{}

Therefore
\begin{align*}
& \Vert S_{\beta} (\tilde{\beta}^{KL} - \beta^*)\Vert^2 \leq \big(\sum_{e\in [e]}\frac{1}{\sigma_e^2}\big)\Big(\Vert\sum_{e\in [E]}\frac{\Sigma_e^{-1}}{\sigma_e^2}\Vert + \sum_{e\in [e]}\frac{\Vert\Sigma_e^{-1}\Vert}{\sigma_e^2}\Big) \sup_{e\in [E]}\Vert \delta_e\Vert^2
\end{align*}
\end{proof}

\subsection{Auxiliary lemma}

For completeness, we rephrase and prove here a simple well-known result that gives a closed-form formula for the KL divergence between two multivariate Gaussians, using the parametrization via the triplet $(\Sigma_e,\beta_e,\sigma_e^2)$. We drop the subscript $X$ from the covariance matrix $\Sigma_{e,X}$ for notation simplicity.

\begin{lemma}\label{lemma:KL-Gaussian}
Consider $\Xb_i \sim \mathcal{N}(\mathbf{0},\Sigma_i)$, $Y_i = \beta_i \Xb_i + \epsilon_i$ where $\epsilon_i \sim \mathcal{N}(0,\sigma_i^2)$ for $i\in \{1,2\}$. Denote by $\pi_i$ the joint distribution of $(\Xb_i, Y_i)$, we have the following identity:
\begin{align*}
    KL(\pi_1 \Vert \pi_2) = & KL\big(\mathcal{N}(\mathbf{0},\Sigma_1)\Vert \mathcal{N}(\mathbf{0},\Sigma_2) \big) + KL(\mathcal{N}(0,\sigma_1^2)\Vert \mathcal{N}(0,\sigma_2^2) +\frac{1}{2}\frac{(\beta_1 - \beta_2)^T \Sigma_1 (\beta_1 - \beta_2)}{\sigma_2^2}
\end{align*}{}
\end{lemma}{}

\begin{proof}\label{proof:lemma:KL-Gaussian}
The distribution $\pi_i$ is a centered multivariate Gaussian distribution characterized by the full covariance matrix $\Sigma_i^f$:
\begin{align*}
    \pi_i = \mathcal{N}\Big(\mathbf{0}; \Sigma^{f}_i \Big) := \mathcal{N}\Bigg( 
    0; 
    \begin{pmatrix}
    \Sigma_i & \Sigma_i \beta_i
    \\ \beta_i^T \Sigma_i & \sigma_i^2 + \beta_i^T\Sigma_i\beta_i
    \end{pmatrix}
    \Bigg)
\end{align*}{}
We have in particular the following identities:
\begin{align*}
    \begin{cases}{}
     \det(\Sigma_i^f) =  \sigma_i^2\det(\Sigma_i)
    \\ (\Sigma_i^f)^{-1} = \begin{pmatrix}{}
    (\Sigma_i^{-1} + \frac{1}{\sigma_i^2}\beta_i\beta_i^T) & -\frac{1}{\sigma_i^2}\beta_i
    \\ -\frac{1}{\sigma_i^2}\beta_i^T & \frac{1}{\sigma_i^2}
    \end{pmatrix}
    \\ Tr\big((\Sigma_2^{f})^{-1}\Sigma_1^{f}\big) = Tr(\Sigma_2^{-1}\Sigma_1) + \frac{\sigma_1^2}{\sigma_2^2} + \frac{(\beta_1 - \beta_2)^T \Sigma_1 (\beta_1 - \beta_2)}{\sigma_2^2}
    \end{cases}
\end{align*}{}
Therefore the KL divergence is given by:
\begin{align*}
    KL( \pi_1\Vert \pi_2) & =\frac{1}{2}\Big( \log\det(\Sigma_2^{f})-\log\det(\Sigma_1^{f})  - (d+1) + Tr\big((\Sigma_2^{f})^{-1}\Sigma_1^{f}\big)\Big)
    \\ &= \frac{1}{2}\Big(\log\det(\Sigma_2)-\log\det(\Sigma_1) +\log(\sigma^2_2)-\log(\sigma_1^2) -d -1 +  Tr(\Sigma_2^{-1}\Sigma_1) 
    \\ & \quad + \frac{\sigma_1^2}{\sigma_2^2}  + \frac{(\beta_1 - \beta_2)^T \Sigma_1 (\beta_1 - \beta_2)}{\sigma_2^2}\Big)
    \\ & = KL\big(\mathcal{N}(\mathbf{0},\Sigma_1)\Vert \mathcal{N}(\mathbf{0},\Sigma_2) \big) + KL(\mathcal{N}(0,\sigma_1^2)\Vert \mathcal{N}(0,\sigma_2^2) +\frac{1}{2}\frac{(\beta_1 - \beta_2)^T \Sigma_1 (\beta_1 - \beta_2)}{\sigma_2^2}
\end{align*}{}
Hence the result.
\end{proof}{}

\end{document}